%% file: sdidStataJournal.tex
\documentclass[varwidth,bib]{statapress}

\usepackage[crop,newcenter,frame]{pagedims}
\usepackage{sj-2}
\usepackage{epsfig}
\usepackage{stata}
\usepackage{shadow}
\sjsetissue{$vv$}{$ii$}{$mm$}{$yyyy$}

\usepackage{amsmath}
\usepackage{bbm}
\usepackage{amsfonts}
\usepackage{booktabs}
\usepackage{hyperref}
\usepackage{setspace}
\usepackage{soul}
\usepackage{lscape}
\usepackage{rotating}
\usepackage{multirow}
\usepackage{floatrow}
\usepackage{rotating,capt-of}
\usepackage{array}
\usepackage[caption=false]{subfig}
\usepackage{bm}
\usepackage{bbm}
\usepackage{url}
\usepackage{color, colortbl}
\definecolor{Gray}{gray}{0.9}

\usepackage{amsthm}
\usepackage[stable]{footmisc}
\usepackage{graphicx}
\usepackage{booktabs}
\usepackage{longtable}
\usepackage{float}
\usepackage{subfloat}
\usepackage{verbatim}
\usepackage{caption}
\usepackage{fancyhdr}

\usepackage{algpseudocode}
\usepackage[ruled,vlined]{algorithm2e}

\usepackage{cleveref}

\hypersetup{
  colorlinks=true,
  linkcolor=blue,
  citecolor=blue,
  filecolor=blue,
  urlcolor=blue
}

        \hypersetup{
           breaklinks=true,
           pdfusetitle=true, 
        }
        
\usepackage{natbib}
\pdfoutput=1

\DeclareMathOperator*{\argmin}{arg\,min}
\begin{document}

\inserttype[]{article}
\author{Susan Athey and Damian Clarke and Guido Imbens and Daniel Pailañir}{%
    Damian Clarke\\Department of Economics \\ University of Chile \\ dclarke@fen.uchile.cl
    \and
    Daniel Pailañir\\Department of Economics \\ University of Chile \\ dpailanir@fen.uchile.cl
    \medskip
    \and
    Susan Athey\\Graduate School of Business\\Stanford University \\ athey@stanford.edu
    \and
    Guido Imbens\\Graduate School of Business \\Stanford University \\ imbens@stanford.edu}
\title[Synthetic Difference In Differences]{Synthetic Difference In Differences Estimation}

\maketitle

\begin{abstract}
  In this paper, we describe a computational implementation of the Synthetic difference-in-differences (SDID) estimator of \citet{Arkhangelskyetal2021} for Stata.  Synthetic difference-in-differences can be used in a wide class of circumstances where treatment effects on some particular policy or event are desired, and repeated observations on treated and untreated units are available over time.  We lay out the theory underlying SDID, both when there is a single treatment adoption date and when adoption is staggered over time, and discuss estimation and inference in each of these cases.  We introduce the \texttt{sdid} command which implements these methods in Stata, and provide a number of examples of use, discussing estimation, inference, and visualization of results.
\end{abstract}
\clearpage
\section{Introduction}
There has been an explosion in recent advances in econometric methods for policy analysis.  A particularly active area is that applied to estimating the impact of exposure to some particular event or policy, when observations are available in a panel or repeated cross section of groups and time (see for example recent surveys by \citet{dCDH2022,Rothetal2022} for reviews of these methods).  A modelling challenge in this setting is in determining what would have happened to exposed units had they been left unexposed.  Should such a counterfactual be estimable from underlying data, causal inference can be conducted by comparing  outcomes in treated units to those in theoretical counterfactual untreated states, under the potential outcome framework (see for example the discussion in \citet{Holand1986,Rubin2005}.  

A substantial number of empirical studies in economics and the social sciences more generally seek to estimate effects in this setting using difference-in-difference (hereafter DID) style designs. Here impacts are inferred by comparing treated to control units, where time-invariant level differences between units are permitted as well as general common trends. However, the drawing of causal inferences requires a parallel trend assumption, which states that in the absence of treatment, treated units would have followed parallel paths to untreated units.  Whether this assumption is reasonable in a particular context is an empirical issue.  Recently, a number of methodologies have sought to loosen this assumption.  This includes procedures in which counterfactual trends can be assumed to deviate from parallel, leading to partial identification \citep{ManskiPepper2018,RambachanRoth2019}, flexible procedures to adequately control for any prevailing differences between treated and control units \citep{BilinskiHatfield2018}, often based on pre-treatment periods only \citep{Bhulleretal2013,GoodmanBacon2021} and IV-style methods which explicitly consider dynamics in pre-treatment periods \citep{Freyaldenhovenetal2019}.

In many cases, parallel trends may be a questionable modelling assumption. One particular solution to the challenge has been the application of synthetic control methods.  Early work in synthetic control explores the setting of comparative case studies, where a single treated unit is observed, and one wishes to construct a matched synthetic control from a larger number of potential donor units \citep{AbadieGardeazabal2003,Abadieetal2010,Abadieetal2015}.  These methods seek to generate a single synthetic control from a unique convex weighting of underlying control units, such that this synthetic control is as closely matched as possible to the treated unit in pre-treatment outcomes, and potentially other covariates.  This weights are optimally generated and fixed over time, potentially assigning zero weight to certain control units, and larger weights to others. This has attracted considerable attention in both empirical applications and theoretical extensions, with recent advances including debiasing procedures \citep{BenMichealetal2021} which can additionally house multiple treatment units \citep{AbadieLHour2021}, more flexible weighting schemes, or constant fixed differences between treated and synthetic control units \citep{DoudchenkoImbens2016,FermanPinto2021}.

A recent particularly flexible modelling option which can be applied in panel data settings seeks to bridge the DID and synthetic control (SC) procedures.  \citet{Arkhangelskyetal2021} propose the Synthetic Difference-in-Differences estimator (SDID), which brings in strengths from both the DID and SC methods.  Like DID models, SDID allows for treated and control units to be trending on entirely different levels prior to a reform of interest.  And like SC methods, SDID seeks to optimally generate a matched control unit which considerably loosens the need for parallel trend assumptions.  Correspondingly, SDID avoids common pitfalls in standard DID and SC methods -- namely an inability to estimate causal relationships if parallel trends are not met in aggregate data in the case of DID, and a requirement that the treated unit be housed within a ``convex hull'' of control units in the case of SC.  \citet{Arkhangelskyetal2021} propose estimation and inference procedures, formally proving consistency and asymptotic normality of the proposed estimator.  What's more, the authors briefly discuss a number of important applied points such as how their estimator can incorporate covariates, and how their estimator can be applied to both multiple treatment units, and even multiple treatment units which adopt treatment in different time periods.

In this paper we describe the {\tt sdid} command (available for download as \citet{PailanirClarke2022}) which implements the SDID estimator in Stata.  This command allows for the simple implementation of the SDID estimator provided that a panel or repeated cross section of data is available covering groups and time periods, and which is strongly balanced.  The command, written principally in Mata, seamlessly incorporates cases where there is a single treated unit, multiple treatment units, and multiple treatment periods. It reports treatment effects laid out in \citet{Arkhangelskyetal2021}, additionally implementing their proposed bootstrap, jackknife and placebo inference procedures.  A number of graphical output options are provided to examine the generation of the SDID estimator and the underlying optimal weight matrices.  While principally written to conduct SDID estimation, the {\tt sdid} command (and the SDID method) nests as possible estimation procedures SC and DID, which can be easily generated to allow comparison of estimation procedures and estimates.\footnote{Code from the original paper was provided in R \citep{HirshbergND}, which can do many of the procedures which {\tt sdid} implements, and indeed, abstracting from differences in pseudo-random number generation, give identical results in cases where similar procedures are possible. A number of useful extensions are available in {\tt sdid}, such as the implementation of estimates in cases where treatment occurs in multiple periods, and alternative manners to include covariates.}

In introducing the command, we first provide a primer on the core methodological points of SDID (as well as comparisons to DID and SC), and then describe how these procedures extend to a setting where treatment adoption occurs over multiple time periods.  We then lay out the command syntax of {\tt sdid}, as well as the elements which are returned to the user.  We provide a number of examples to illustrate the use of the SDID method in Stata, both based upon a well-known example of California's passage of Proposition 99, an anti-smoking measure previously presented in \citet{Abadieetal2010,Arkhangelskyetal2021} in which a single state adopts a treatment at a given time, as well as an example where exposure to a policy occurs at mutiple periods: the case of parliamentary gender quotas studied by \citet{Bhalotraetal2022}.  We conclude by making a number of practical points on the computational implementation of this estimator.

\section{Methods}
\label{scn:methods}
\subsection{The Canonical Synthetic Difference-in-Differences Procedure}
The synthetic DID procedure, hereafter SDID, is developed in \citet{Arkhangelskyetal2021}, and we lay out principal details here.  As input, SDID requires a balanced panel of $N$ units or groups, observed over $T$ time periods.  An outcome, denoted $Y_{it}$, is observed for each unit $i$ in each period $t$. Some, but not all, of these observations are treated with a specific binary variable of interest, denoted $W_{it}$.  This treatment variable $W_{it}=1$ if observation $i$ is treated by time $t$, otherwise, $W_{it}=0$ indicates that unit $i$ is untreated at time $t$.  Here, we assume that there is a single adoption period for treated units, which \citet{Arkhangelskyetal2021} refer to as a `block treatment assignment'.  In section \ref{sscn:SAD}, we extend this to a `staggered adoption design' \citep{AtheyImbens2022}, where treated units adopt treatment at varying points.  A key element of both of these designs is that once treated, units are assumed to remain exposed to treatment forever thereafter.  In the particular setting of SDID, no always treated units can be included in estimation.  For estimation to proceed, we require at least two pre-treatment periods off of which to determine control units.

The goal of SDID is to consistently estimate the causal effect of receipt of policy or treatment $W_{it}$, (an average treatment effect on the treated, or ATT) even if we do not believe in the parallel trends assumption between all treatment and control units on average.  Estimation of the ATT proceeds as follows: 
\begin{equation}
    \label{eqn:sdid}
    \left(\widehat\tau^{sdid},\widehat\mu,\widehat\alpha,\widehat\beta\right)=\argmin_{\tau,\mu,\alpha,\beta}\left\{\sum_{i=1}^N\sum_{t=1}^T(Y_{it}-\mu-\alpha_i-\beta_t-W_{it}\tau)^2\widehat\omega^{sdid}_i\widehat\lambda^{sdid}_t \right\}
\end{equation}
where the estimand is the ATT, generated from a two-way fixed effect regression, with optimally chosen weights $\widehat\omega^{sdid}_i$ and $\widehat\lambda^{sdid}_t$ discussed below.  Note that here, this procedure flexibly allows for shared temporal aggregate factors given the estimation of time fixed effects $\beta_t$ and time invariant unit-specific factors given the estimation of unit fixed effects $\alpha_i$.  As is standard in fully saturated fixed-effect models, one $\alpha_i$ and one $\beta_t$ fixed effect are normalized to zero to avoid multi-colinearity.  The presence of unit-fixed effects implies that SDID will simply seek to match treated and control units on pre-treatment trends, and \emph{not} necessarily on both pre-treatment trends and levels, allowing for a constant difference between treatment and control units.

In this setting, it is illustrative to consider how the SDID procedure compares to the traditional synthetic control method of \citet{Abadieetal2010}, as well as the baseline DID procedure.  The standard DID procedure consists of precisely the same two-way fixed effect OLS procedure, simply assigning equal weights to all time periods and groups:
\begin{equation}
    \label{eqn:did}
    \left(\widehat\tau^{did},\widehat\mu,\widehat\alpha,\widehat\beta\right)=\argmin_{\tau,\mu,\alpha,\beta}\left\{\sum_{i=1}^N\sum_{t=1}^T(Y_{it}-\mu-\alpha_i-\beta_t-W_{it}\tau)^2 \right\}.
\end{equation}
The synthetic control, on the other hand, maintains optimally chosen unit-specific weights $\omega$ (as laid out below), however does not seek to optimally consider time periods via time weights, and omits unit fixed effects $\alpha_i$ implying that the synthetic control and treated units should maintain approximately equivalent pre-treatment levels, as well as trends.
\begin{equation}
    \label{eqn:sc}
    \left(\widehat\tau^{sc},\widehat\mu,\widehat\beta\right)=\argmin_{\tau,\mu,\beta}\left\{\sum_{i=1}^N\sum_{t=1}^T(Y_{it}-\mu-\beta_t-W_{it}\tau)^2\widehat\omega^{sc}_i \right\}
\end{equation}
From (\ref{eqn:did})-(\ref{eqn:sc}) it is clear that the SDID procedure offers greater flexibility than both the DID and SC procedures; in the case of DID by permitting a violation of parallel trends in aggregate data, and in the case of SC, by both additionally seeking to optimally weight time periods when considering counterfactual outcomes, and allowing for level differences between treatment and control groups.

The selection of unit weights, $\omega$, as inputs to (\ref{eqn:sdid}) (and (\ref{eqn:sc})) seeks to ensure that comparison is made between treated units and controls which were approximately following parallel trends prior to the adoption of treatments.  The selection of time weights, $\lambda$ in the case of SDID seeks to draw more weight from pre-treatment periods which are more similar to post-treatment periods, in the sense of finding a constant difference between each control unit's post treatment average, and pre-treatment weighted averages across all selected controls.  Specifically, as laid out in \citet{Arkhangelskyetal2021}, unit-specific weights are found by resolving:  
\begin{eqnarray}
\left(\widehat\omega_0,\widehat\omega^{sdid}\right)&=&\argmin_{\omega_0\in\mathbb{R},\omega\in\Omega}\sum_{t=1}^{T_{pre}}\left(\omega_0 + \sum_{i=1}^{N_{co}}\omega_iY_{it}-\frac{1}{N_{tr}}\sum_{i=N_{co}+1}^N Y_{it}\right)^2 + \zeta^2 T_{pre}||\omega||^2_2 
\end{eqnarray}
\[ \text{ where } \Omega = \left\{ \omega\in\mathbb{R}^N_{+}, \text{ with } \sum_{i=1}^{N_{co}}\omega_i=1 \text{ and } \omega_i = \frac{1}{N_{tr}} \text{ for all } i=N_{co}+1,\ldots, N \right\}, \]
$||\omega||_2$ refers to the Euclidean norm and $\zeta$ is a regularization parameter laid out in \citet[pp.\ 4091-4092]{Arkhangelskyetal2021}.\footnote{For the sake of completion, this regularization parameter is calculated as $\zeta = (N_{tr}\times T_{post})^{1/4}\widehat\sigma,$ where:
\[\widehat\sigma^2=\frac{1}{N_{co}(T_{pre}-1)}\sum_{i=1}^{N_{co}}\sum_{t=1}^{T_{pre}-1}(\Delta_{it}-\bar{\Delta})^2, \ \Delta_{it}=Y_{i,(t+1)}-Y_{it},  \text{ and } \bar{\Delta}=\frac{1}{N_{co}(T_{pre}-1)}\sum_{i=1}^{N_{co}}\sum_{t=1}^{T_{pre}-1}\Delta_{it}.
\]
\label{foot:zeta}}
This leads to a vector of $N_{co}$ non-negative weights plus an intercept $\omega_0$.  The weights $\omega_i$ for all $i\in\{1,\ldots,N_{co}\}$ imply that absolute difference between control and treatment trends units should be minimized over all pre-treatment periods, while $\omega_0$ initially allows for a constant difference between treatment and controls over time.  Together, these imply that units will follow parallel pre-trends, though provided $\omega_0\neq0$, not \emph{identical} pre-trends.

In the case of time weights, a similar procedure is followed, finding weights which minimize the following objective function:
\begin{eqnarray}
\label{eqn:lambda}
\left(\widehat\lambda_0,\widehat\lambda^{sdid}\right)&=&\argmin_{\lambda_0\in\mathbb{R},\lambda\in\Lambda}\sum_{i=1}^{N_{co}}\left(\lambda_0 + \sum_{t=1}^{T_{pre}}\lambda_tY_{it}-\frac{1}{T_{post}}\sum_{t=T_{pre}+1}^T Y_{it}\right)^2 + \zeta^2 N_{co}||\lambda||^2 
\end{eqnarray}
\[ \text{ where } \Lambda = \left\{ \lambda\in\mathbb{R}^T_{+}, \text{ with } \sum_{t=1}^{T_{pre}}\lambda_t=1 \text{ and } \lambda_t = \frac{1}{T_{post}} \text{ for all } t=T_{pre}+1,\ldots, T \right\}, \]
where the final term in (\ref{eqn:lambda}) is a very small regularization term to ensure uniqueness of time weights, where $\zeta=1\times10^{-6}\widehat\sigma$, and $\widehat\sigma$ is defined as in \cref{foot:zeta}.

This estimation procedure is summarized in \citet[algorithm 1]{Arkhangelskyetal2021}, reproduced in Appendix \ref{app:algorithms} here for ease of access.  \citet{Arkhangelskyetal2021} also prove that the estimator is asymptotically normal, suggesting that confidence intervals on $\tau$ can be constructed as:
\[
\widehat{\tau}^{sdid}\pm z_{\alpha/2}\sqrt{\widehat{V}_\tau},
\]
where $z_{\alpha/2}$ refers to the inverse normal density function at percentile $\alpha/2$ should one wish to compute 1-$\alpha$ confidence intervals.  These confidence intervals thus simply require an estimate of the variance of $\tau$, $\widehat{V}_{\tau}$.  \citet{Arkhangelskyetal2021} propose three specific procedures to estimate this variance: a block bootstrap, a jackknife, or a permutation-based approach.

The block (also known as clustered) bootstrap approach, consists of taking some large number, $B$, of bootstrap resamples over units, where units $i$ are the resampled blocks in the block bootstrap procedure. Provided that a given resample does not consist entirely of treated, or entirely of control units, the quantity $\widehat\tau^{sdid}$ is re-estimated, and denoted as $\widehat\tau_{(b)}^{sdid}$ for each bootstrap resample.  The bootstrap variance $\widehat{V}^{(b)}_\tau$ is then calculated  as the variance of resampled estimates $\widehat\tau_{(b)}^{sdid}$ across all $B$ resamples.  The bootstrap algorithm is defined in \citet[algorithm 2]{Arkhangelskyetal2021}, reproduced here in appendix \ref{app:algorithms}.  This bootstrap procedure is observed in simulation to have particularly good properties, but has two particular drawbacks, justifying alternative inference procedures.  The first is that it may be computationally costly, given that in each bootstrap resample the entire synthetic DID procedure is re-estimated, including the estimation of optimal weights.  This is especially computationally expensive in cases where working with large samples, or where covariates are included, as discussed at more length below.  The second, is that formal proofs of asymptotic normality rely on the number of treated units being large, and as such, estimated variance, and confidence intervals, may be unreliable when the number of treated units is small.  

An alternative estimator which significantly reduces the computational burden inherent in the bootstrap is estimating a jackknife variance for $\widehat\tau^{sdid}$.  This procedure consists of iterating over all units in the data, in each iteration removing the given unit, and recalculating $\widehat\tau^{sdid}$, denoted $\widehat\tau_{(-i)}^{sdid}$, \emph{maintaining fixed} the optimal weights for $\omega$ and $\lambda$ calculated in the original SDID estimate.  The jackknife variance, $\widehat{V}^{(jack)}_\tau$ is then calculated based on the variance of all $\widehat\tau_{(-i)}^{sdid}$ estimates, following \citet[Algorithm 3]{Arkhangelskyetal2021} (refer to Appendix \ref{app:algorithms} here). In this case, each iteration saves on recalculating optimal weights, and as documented by \citet{Arkhangelskyetal2021}, provide a variance leading to conservative confidence intervals, without the computational burden imposed by the bootstrap.  Once again, asymptotic normality relies on there being a large number of treated units, and in particular if only 1 treated unit is observed -- as is often the case in comparative case studies -- the jackknife will not even be defined given that a $\widehat\tau_{(-i)}^{sdid}$ term will be undefined when removing the single treated unit.

Given limits to these inference options when the number of treated units is small, an alternative placebo, or permutation-based, inference procedure is proposed.  This consists of, first, conserving just the control units, and then randomly assigning the same treatment structure to these control units, as a placebo treatment.  Based on this placebo treatment, we then re-estimate $\widehat\tau^{sdid}$, denoted $\widehat\tau_{(p)}^{sdid}$.  This procedure is repeated many times, giving rise to a vector of estimates $\widehat\tau_{(p)}^{sdid}$, and the placebo variance, $\widehat{V}^{(placebo)}_\tau$, can be estimated as the variance of this vector.  This is formally defined in \citet[algorithm 4]{Arkhangelskyetal2021}, and appendix \ref{app:algorithms} here.  It is important to note that in the case of this placebo-based variance, homoskedasticity across units is required, given that the variance is based off placebo assignments of treatment made only within the control group.

\subsection{Conditioning on Covariates}
\label{sscn:controls}
So far,	we have	limited	exposition to cases where one wishes to study outcomes $Y_{it}$, and their evolution in treated and synthetic control units.  However, in certain settings, it may be	of relevance to	condition on exogenous time-varying covariates $X_{it}$.  \citet{Arkhangelskyetal2021} note that in this case, we can proceed by applying the SDID algorithm to the residuals calculated as:
\begin{equation}
    \label{eqn:covariates}
    Y_{it}^{res} = Y_{it}-X_{it}\widehat\beta,
\end{equation}
where $\widehat\beta$ comes from regression of $Y_{it}$ on $X_{it}$. This procedure, in which the synthetic DID process will be applied to the residuals $Y_{it}^{res}$, is different to the logic of synthetic controls following \citet{Abadieetal2010}. In \citeauthor{Abadieetal2010}'s conception, when covariates are included the synthetic control is chosen to ensure that these covariates are as closely matched as possible between treated and synthetic control units.  However in the SDID conception, covariate adjustment is viewed as a pre-processing task, which removes the impact of changes in covariates from the outcome $Y_{it}$ prior to calculating the synthetic control.  Along with their paper, \citet{Arkhangelskyetal2021} provide an implementation of their algorithm in R \citep{HirshbergND}, and in practice they condition out these variables $X_{it}$ by finding $\widehat\beta$ within an optimization procedure which additionally allows for the efficient calculation of optimal weights $\omega$ and $\lambda$.  In the \texttt{sdid} code described below, we follow \citet{HirshbergND} in implementing this efficient optimization procedure (the \citet{FrankWolfe1956} solver), however there are a number of potential complications which can arise in this manner of dealing with covariates, and as such, alternative procedures are also available.

A first potential issue is purely numerical.  In the Frank-Wolfe solver discussed above, a minimum point is assumed to be found when successive iterations of the solver lead to arbitrarily small changes in all parameters estimated.\footnote{In the case of convex functions such as that in (\ref{eqn:sdid}), the Frank-Wolfe solver finds a global minima, see for example \citet{Lawphongpanich2009}.}  Where these parameters include coefficients on covariates, in extreme cases the solution found for (\ref{eqn:sdid}) can be sensitive to the scaling of covariates.  This occurs in particular when covariates have very large magnitudes and variances.  In such cases, the inclusion of covariates in (\ref{eqn:sdid}) can cause optimization routines to suggest solutions which are not actually globally optimal, given that successive movements in $\widehat\beta$ can be very small.  In extreme cases, this can imply that when multiplying all variables $X_{it}$ by a large constant values, the estimated treatment effect can vary.  While this issue can be addressed by using smaller tolerances for defining stopping rules in the optimization procedure, it can be addressed in a more simple way if all covariates are first re-standardized as Z-scores, implying that no very-high-variance variables are included, while capturing the same underlying variation in covariates.  


A second, potentially more complicated point is described by \citet{Kranz2022}.  He notes that in certain settings, specifically where the relationship between covariates and the outcome vary over time differentially in treatment and control groups, the procedure described above may fail to capture the true relationship between covariates and the outcome of interest, and may subsequently lead to bias in estimated treatment effects.  He proposes a slightly altered methodology of controlling for covariates.  Namely, his suggestion is to first estimate a two-way fixed effect regression of the dependent variable on covariates (plus time and unit fixed effects), using \emph{only} observations which are not exposed, or not exposed yet, to treatment, i.e.\ sub-setting to observations for which $W_{it}=0$.  Based on the regression
$Y_{it} = X_{it}\beta + \mu_i + \lambda_t + \varepsilon_{it}$
for units with $W_{it}=0$,
coefficients $\widehat\beta$ estimated by OLS can be used to follow the procedure in (\ref{eqn:covariates}), and SDID can then be conducted.  An additional benefit of this method is that unlike the optimized method described previously, $\widehat\beta$ is calculated in a single step via OLS rather than an interative optimization procedure, which can lead to substantial speed ups in computation time.  

We document these methods based on a number of empirical example in the following section.  It is worth noting that regardless of which procedure one uses, the implementation of SDID follows the suggestion laid out in \citet[footnote 4]{Arkhangelskyetal2021} and (\ref{eqn:covariates}) above.  What varies between the former and the latter procedures discussed here is the manner with which one estimates coefficients on covariates, $\widehat\beta$.

\subsection{The Staggered Adoption Design}
\label{sscn:SAD}
The design discussed up to this point assumes block assignment, or that all units are either controls, or are treated in a single unit of time.  However, in \citet[Appendix A]{Arkhangelskyetal2021}, they note that this procedure can be extended to a staggered adoption design, where treated units adopt treatment at varying moments of time. Here we lay out the formal details related to the staggered adoption design, focusing first on estimation of an aggregate treatment effect, and then on extending the three inference procedures laid out previously into a staggered adoption setting.  This proposal is one potential way to deal with staggered adoption settings, though there are other possible manners to proceed -- see for example \citet{BenMichealetal2021}, or \citet[Appendix A]{Arkhangelskyetal2021}.  In cases where the number of pure control units is small, this proposal may not necessarily be very effective given challenges in finding appropriate counterfactuals for each adoption-specific period. 

\paragraph{Estimation} 
Unlike the block assignment case where a single pre- versus post-treatment date can be used to conduct estimation, in the staggered adoption design, multiple adoption dates are observed.  Consider for example the treatment matrix below, consisting of 8 units, 2 of which (1 and 2) are untreated, while the other 6 are treated, however at varying points.  \[
\mathbf{W} =
\begin{pmatrix}
  & t_1 & t_2 & t_3 & t_4 & t_5 & t_6 & t_7 & t_8\\
1 &  0  &  0  &  0  &  0  &  0  &  0  &  0  &  0 \\
2 &  0  &  0  &  0  &  0  &  0  &  0  &  0  &  0 \\
3 &  0  &  0  &  0  &  0  &  0  &  0  &  1  &  1 \\
4 &  0  &  0  &  0  &  0  &  0  &  0  &  1  &  1 \\
5 &  0  &  0  &  0  &  1  &  1  &  1  &  1  &  1 \\
6 &  0  &  0  &  0  &  1  &  1  &  1  &  1  &  1 \\
7 &  0  &  0  &  1  &  1  &  1  &  1  &  1  &  1 \\
8 &  0  &  0  &  1  &  1  &  1  &  1  &  1  &  1 \\
\end{pmatrix}
\]
This single staggered treatment matrix $\mathbf{W}$ can be broken down into adoption date specific matrices, $\mathbf{W}^{1}$, $\mathbf{W}^{2}$ and $\mathbf{W}^{3}$, or generically, $\mathbf{W}^{1}, \ldots, \mathbf{W}^{A}$, where $A$ indicates the number of distinct adoption dates.  Additionally, a row vector $\mathbf{A}$ consisting of $A$ elements contains these distinct adoption periods.  In this specific setting where units first adopt treatment in period $t_3$ (units 7 and 8), $t_4$ (units 5 and 6), and $t_7$ (units 3 and 4), the adoption date vector consists simple of periods 3, 4 and 7.
\[
\mathbf{A} = \begin{pmatrix}
  3 & 4 & 7 
\end{pmatrix}
\]
Finally, adoption-specific matrices $\mathbf{W}^{1}$-$\mathbf{W}^{3}$ simply consist of pure treated units, and units which adopt in this specific period, as below:
\[
\mathbf{W}^1 =
\begin{pmatrix}
  & t_1 & t_2 & t_3 & t_4 & t_5 & t_6 & t_7 & t_8\\
1 &  0  &  0  &  0  &  0  &  0  &  0  &  0  &  0 \\
2 &  0  &  0  &  0  &  0  &  0  &  0  &  0  &  0 \\
3 &  0  &  0  &  0  &  0  &  0  &  0  &  1  &  1 \\
4 &  0  &  0  &  0  &  0  &  0  &  0  &  1  &  1 \\
\end{pmatrix}, 
\]
\[
\mathbf{W}^2 =
\begin{pmatrix}
  & t_1 & t_2 & t_3 & t_4 & t_5 & t_6 & t_7 & t_8\\
1 &  0  &  0  &  0  &  0  &  0  &  0  &  0  &  0 \\
2 &  0  &  0  &  0  &  0  &  0  &  0  &  0  &  0 \\
5 &  0  &  0  &  0  &  1  &  1  &  1  &  1  &  1 \\
6 &  0  &  0  &  0  &  1  &  1  &  1  &  1  &  1 \\
\end{pmatrix},
\]
\[
\mathbf{W}^3 =
\begin{pmatrix}
  & t_1 & t_2 & t_3 & t_4 & t_5 & t_6 & t_7 & t_8\\
1 &  0  &  0  &  0  &  0  &  0  &  0  &  0  &  0 \\
2 &  0  &  0  &  0  &  0  &  0  &  0  &  0  &  0 \\
7 &  0  &  0  &  1  &  1  &  1  &  1  &  1  &  1 \\
8 &  0  &  0  &  1  &  1  &  1  &  1  &  1  &  1 \\
\end{pmatrix}.
\]

As laid out in \citet[Appendix A]{Arkhangelskyetal2021}, the average treatment effect on the treated can then be calculated by applying the synthetic DID estimator to each of these 3 adoption-specific samples, and calculating a weighted average of the three estimators, where weights are assigned based on the relative number of treated units and time periods in each adoption group.  Generically, this ATT is calculated based on adoption-specific SDID estimates as:
\begin{equation}
\label{eqn:ATT}
\widehat{ATT}=\sum_{\text{for } a \in \mathbf{A}}\frac{T_{post}^a}{T_{post}}\times\hat{\tau}^{sdid}_{a}
\end{equation}
where $T_{post}$ refers to the total number of post-treatment periods observed in treated units.  This estimation procedures is laid out formally in Algorithm \ref{alg:StaggeredEst} below. 

Note that in this case, while the parameter interest is likely the treatment effect $ATT$ or adoption specific $\tau^{sdid}_a$ parameters, each adoption period is associated with an optimal unit and time weight vector $\omega^{sdid}_a$ and $\lambda^{sdid}_a$, which can be returned following estimation.

\begin{algorithm}[H]
\caption{Estimation of the ATT with staggered adoption}
\label{alg:StaggeredEst}
Data: \textbf{Y}, \textbf{W}, \textbf{A}. \\
Result: Point estimate $\widehat{ATT}$ and adoption-specific values $\hat{\tau}^{sdid}_a$, $\hat\omega^{sdid}_a$ and $\hat\lambda^{sdid}_a$ for all a $\in \mathbf{A}$.  \\ \ \\
\For{a $\in \mathbf{A}$}
{
1. Subset \textbf{Y} and \textbf{W} to units who are pure controls, and who first adopt treatment in period $t=a$. \;
2. Compute regularization parameter $\zeta$ \;
3. Compute unit weights $\hat\omega^{sdid}_a$ \;
4. Compute time weights $\hat\lambda^{sdid}_a$ \;
5. Compute the SDID estimator via the weighted DID regression \;
\[\left(\hat\tau^{sdid}_a,\hat\mu_a,\hat\alpha_a,\hat\beta_a\right)=\argmin_{\tau,\mu,\alpha,\beta}\left\{\sum_{i=1}^N\sum_{t=1}^T(Y_{it}-\mu-\alpha_i-\beta_t-W_{it}\tau)^2\widehat\omega^{sdid}_{a,i}\hat\lambda^{sdid}_{a,t} \right\}\]
}\textbf{end}\\ \ \\
6. Compute ATT across adoption-specific SDID estimates \\
\[\widehat{ATT}=\sum_{\text{for } a\in \mathbf{A}}\frac{T_{post}^a}{T_{post}}\times\hat{\tau}^{sdid}_{a}\] \\
\end{algorithm}

\paragraph{Inference}
In the staggered adoption design, estimated treatment effects are simply a multi-period extension of the underlying SDID algorithm, in each case working with the relevant pure control and treated sub-sample.  Thus, inference can be conducted in the staggered adoption design under similar resample or placebo procedures.  Here we discuss inference following each of the bootstrap, jackknife, or placebo procedures laid out in \citet{Arkhangelskyetal2021}, applied to a staggered adoption setting.  We note that in this design, it is likely the case that one wishes to conduct inference on the treatment effect, ${ATT}$ from (\ref{eqn:ATT}).  Thus in the below, we propose inference details for this estimand, additionally noting that standard errors and confidence intervals on adoption-specific SDID parameters $\tau^{sdid}_a$ come built-in as part of these procedures.

Consider first the case of bootstrap inference.  Suppose that one wishes to estimate standard errors or generate confidence intervals on the global treatment effect $ATT$.  A bootstrap procedure can be conducted based on many clustered bootstrap resamples over the entire initial dataset, where in each case, a resampled ATT estimate $\widehat{ATT}^{b}$ is generated, following Algorithm \ref{alg:StaggeredEst}.  Based on many such resampled estimates, the bootstrap variance can be calculated as the variance of these resamples.  We lay out the bootstrap variance estimate below in Algorithm \ref{alg:BoostrapStaggered}.
\begin{algorithm}
\caption{Bootstrap inference in the staggered adoption setting}
\label{alg:BoostrapStaggered}
Data: \textbf{Y}, \textbf{W}, \textbf{A}, $B$. \\
Result: Variance estimator $\widehat{V}_{ATT}^{cb}$.  Additionally, variance for each adoption specific estimate $\widehat{V}_{\tau_a}^{cb}$ for all $a\in\mathbf{A}$. \\ \ \\
\For{b $\leftarrow$ 1 to B}
{
1. Construct a bootstrap dataset $(\mathbf{Y}^{(b)},\mathbf{W}^{(b)},\mathbf{A}^{(b)})$ by sampling $N$ rows of $(\mathbf{Y},\mathbf{W})$ with replacement, and generating $\mathbf{A}$ as the corresponding adoption vector \;
2. \If{the bootstrap sample has no treated units or no control units} 
{
\hspace{0.2cm}Discard resample and go to 1 \;
}\hspace{0.4cm}\textbf{end}\\ 
3. Compute SDID estimate $ATT^{(b)}$ following Algorithm \ref{alg:StaggeredEst} based on $(\mathbf{Y}^{(b)},\mathbf{W}^{(b)},\mathbf{A}^{(b)})$.  Similarly, generate a vector of adoption-date specific resampled SDID estimates $\tau_{a}^{(b)}$ for all $a\in\mathbf{A}^{(b)}$ \;
}\textbf{end}\\ \ \\
4. Define $\widehat{V}_{ATT}^{cb}=\frac{1}{B}\sum_{b=1}^B\left(\widehat{ATT^{(b)}}-\frac{1}{B}\sum_{b=1}^B\widehat{ATT^{(b)}}\right)^2$.  Similarly, estimate adoption-date specific variances for each $\tau_{a}^{sdid}$ estimate as the variance over each $\tau_{a}^{(b)}$;
\end{algorithm}
Note that as in the block treatment design, this bootstrap procedure requires the number of treated units to grow with $N$ within each adoption period.  As such, if a very small number of treated units exist for certain adoption periods, placebo inference is likely preferable.  Similarly, as laid out in the block treatment design, the bootstrap procedure re-estimates optimal weight matrices in each resample, and can be computationally expensive in cases where samples are large.  

An alternative inference procedure, which is less computationally intensive but similarly based on asymptotic arguments with a large number of states, and many treated units, is based on the jackknife.  Here, optimal weight matrices calculated for each adoption-specific estimate $\tau^{sdid}_a$ in Algorithm \ref{alg:StaggeredEst} are treated as fixed, and provided as inputs to a jackknife procedure, described below in Algorithm \ref{alg:jackknifeStaggered}.  Below, these matrices, which consist of weights for each adoption period $a\in\mathbf{A}$, are denoted as $\bm{\omega}$, $\bm{\lambda}$.\footnote{In the case of $\bm{\omega}$, this is an $A\times N_{co}$ matrix, where for each row $a$, the weight assigned to each control unit in this particular adoption period is recorded.  In the case of $\bm{\lambda}$, this is a matrix containing $A$ columns, where each column consists of as many rows as the number of pre-treatment periods to this specific adoption date.  In each cell, the weight assigned to a particular pre-treatment year for each adoption period is recorded.}  Note that in Algorithm \ref{alg:jackknifeStaggered}, notation $(-i)$ refers to a standard jackknife estimator, removing a single state ($i$) in each iteration.  In cases where $i$ refers to a treated unit, the ATT will be calculated removing this particular treated unit.  For this reason, the jackknife estimator will not be defined in cases where any single adoption period has only one treated unit, as in this case, $\widehat\tau_{a}^{(-i)}$ will not be defined.
\begin{algorithm}
\caption{Jackknife inference in the staggered adoption setting}
\label{alg:jackknifeStaggered}
Inputs: \textbf{Y}, \textbf{W}, \textbf{A}, $\widehat{\bm{\omega}}$, $\widehat{\bm{\lambda}}$, $\widehat{ATT}$. \\
Result: Variance estimator $\widehat{V}_{ATT}$ \\ \ \\
\For{i $\leftarrow$ 1 to N}
{
1. Compute SDID estimate $ATT^{(-i)}$ following Algorithm \ref{alg:StaggeredEst} based on $(\mathbf{Y}^{(-i)},\mathbf{W}^{(-i)},\mathbf{A}^{})$.  Similarly, generate a vector of adoption-date specific resampled SDID estimates $\tau_{a}^{(-i)}$ for all $a\in\mathbf{A}$\;
}\textbf{end}\\ \ \\
2. Compute $\widehat{V}_{ATT}^{jack}=(N-1)N^{-1}\sum_{i=1}^N\left(\widehat{ATT}^{(-i)}-\widehat{ATT}\right)^2$;
\end{algorithm}

Finally, in cases where the number of treated units is small, and concerns related to the validity of the previous variance estimators exists, the placebo inference procedure defined in algorithm \ref{alg:placeboStaggered} can be used.  Here, this is defined for the staggered adoption case, generalising Algorithm 4 of \citet{Arkhangelskyetal2021}.  To conduct this procedure, placebo treatments are randomly assigned based on the actual treatment structure, however \emph{only} to the control units. Based on these placebo assignments, placebo values for $ATT$ are generated, which can be used to calculate the variance as laid out in Algorithm \ref{alg:placeboStaggered}.  It is important to note that such a procedure will only be feasible in cases where the number of control units is strictly larger than the number of treated units (or hence placebo assignments will not be feasible), and, as laid out in \citet{Arkhangelskyetal2021,ConleyTaber2011}, such a procedure relies on homoskedasticity across units, as otherwise the variance of the treatment effect on the treated could not be inferred from variation in assignment of placebo treatments to control units.

\begin{algorithm}
\caption{Placebo inference in the staggered adoption}
\label{alg:placeboStaggered}
Inputs: $\mathbf{Y}_{co}$, $N_{tr}$, $B$. \\
Result: Variance estimator $\widehat{V}_{ATT}^{placebo}$ \\ \ \\
\For{b $\leftarrow$ 1 to B}
{
1. Sample $N_{tr}$ out of the $N_{co}$ control units without replacment to `receive the placebo' \;
2. Construct a placebo treatment matrix $\mathbf{W}^{(b)}_{co}$, for the controls \; 
3. Compute SDID estimate $ATT^{(b)}$ following algorithm \ref{alg:StaggeredEst} based on $(\mathbf{Y}_{co},\mathbf{W}^{(b)}_{co},\mathbf{A}^{(b)})$ \;
}\textbf{end}\\ \ \\
4. Define $\widehat{V}_{ATT}^{placebo}=\frac{1}{B}\sum_{b=1}^B\left(\widehat{ATT^{(b)}}-\frac{1}{B}\sum_{b=1}^B\widehat{ATT^{(b)}}\right)^2$;
\end{algorithm}


\section{The sdid command}
\label{scn:syntax}
Synthetic Difference-in-Differences can be implemented in Stata using the following command syntax:

\begin{stsyntax}
  sdid {\it depvar groupvar timevar treatment \/}
  \optif\
  \optin\
  ,
  vce({\it type})
	\optional{
	covariates({\it varlist, type})
	seed({\it \#\/})
	reps({\it \#\/})
	method({\it type})
    graph
    g1on
	g1\_opt({\it string\/})
	g2\_opt({\it string\/})
	graph\_export({\it string, type\/})
	msize({\it markersizestyle})
    unstandardized
    mattitles
	}
\end{stsyntax}

where \emph{depvar} describes the dependent variable in a balanced panel of units (\emph{groupvar}) and time periods (\emph{timevar}).  The variable which indicates units which are treated at each time period, which accumulates over time, is indicated as \emph{treatment}.  Note that here, it is not necessary for users to specify whether the design is a block or staggered adoption design, as this will be inferred based off the data structure.  Optionally, \emph{if} and \emph{in} can be specified, provided that this does not result in imbalance in the panel.  Required and permitted options are discussed below, followed by a description of objects returned by \texttt{sdid}.

\subsubsection{Options}

\noindent
\leftskip 0.1in
\parindent -0.1in

\texttt{vce({\it type})} is a required option. This must be one of bootstrap, jackknife, placebo or noinference, where in each case inference proceeds following the specified method.  In the case of bootstrap, this is only permitted if greater than one unit is treated.  In the case of jackknife, this is only permitted if greater than one unit is treated in each treatment period (if multiple treatment periods are considered).  In the case of placebo, this requires at least one more control than treated unit to allow for permutations to be constructed.  In each case, inference follows the specific algorithm laid out in \citet{Arkhangelskyetal2021}. We allow the no inference option (noinference) should one wish to simply generate the point estimator.  This is useful if you wish to plot outcome trends without the added computational time associated with inference procedures.

\texttt{covariates({\it varlist, type})} Covariates should be included as a varlist, and if specified, treatment and control units will be adjusted based on covariates in the synthetic difference-in-differences procedure.  Optionally, type may be specified, which indicates how covariate adjustment will occur.  If the type is indicated as ``optimized'' (the default) this will follow the method described in \citet{Arkhangelskyetal2021}, footnote 4, where SDID is applied to the residuals of all units after regression adjustment.  However, this has been observed to be problematic at times (refer to \citet{Kranz2022}), and is also sensitive to optimization if covariates have high dispersion.  Thus, an alternative type is implemented (``projected''), which consists of conducting regression adjustment based on parameters estimated only in untreated units.  This type follows the procedure proposed by \citet{Kranz2022} (xsynth in R), and is observed to be more stable in some implementations (and at times, considerably faster). {\tt sdid} will run simple checks on the covariates indicated and return an error if covariates are constant, to avoid multicolineality.  However, prior to running sdid, you are encouraged to ensure that covariates are not perfectly multicolinear with other covariates and state and year fixed effects, in a simple two-way fixed effect regression.  If perfectly multi-colinear covariates are included sdid will execute without errors, however where type is ``optimized'', the procedure may be sensitive to the inclusion of redundant covariates.

\texttt{seed({\it \#\/})} Define the seed for pseudo-random numbers.

\texttt{reps({\it \#\/})} Set the number of repetitions used in the calculation of bootstrap and placebo standard errors. Default is 50 repetitions.  Larger values should be preferred where possible.

\texttt{method({\it type})} this option allows you to change the estimation method. The type must be one of sdid, did or sc, where sdid refers to synthetic difference-in-differences, sc refers to synthetic control, and did refers to difference-in-differences. By default, sdid is enabled.

\texttt{graph} if this option is specified, graphs will be displayed showing unit and time weights as well as outcome trends as per Figure 1 from \citet{Arkhangelskyetal2021}.

\texttt{g1on} this option activates the unit-specific weight graph. By default g1 is off.

\texttt{g1\_opt({\it string\/})} option to modify the appearance of the unit-specific weight graph.  These options adjust the underlying scatter plot, so should be consistent with twoway scatter plots.

\texttt{g2\_opt({\it string\/})} option to modify the appearance of the outcome trend graphs.  These options adjust the underlying line plot, so should be consistent with twoway line plots.

\texttt{graph\_export({\it string, type\/})} Graphs will be saved as weightsYYYY and trendsYYYY for each of the unit-specific weights and outcome trends respectively, where YYYY refers to each treatment adoption period.  Two graphs will be generated for each treatment adoption period provided that \texttt{g1on} is specified, otherwise a single graph will be generated for each adoption period. If this option is specified, type must be specified, which refers to a valid Stata graph type (eg ``.eps'', ``.pdf'', and so forth). Optionally, a stub can be specified, in which case this will be prepended to exported graph names.

\texttt{msize({\it markersizestyle})} allows you to modify the size of the marker for graph 1.

\texttt{unstandardized} if controls are included and the ``optimized'' method is specified, controls will be standardized as Z-scores prior to finding optimal weights. This avoids problems with optimization when control variables have very high dispersion. If unstandardized is specified, controls will simply be entered in their original units. This option should be used with care.

\texttt{mattitles} requests labels to be added to the returned {\tt e(omega)} weight matrix providing names (in string) for the unit variables which generate the synthetic control group in each case.  If mattitles is not indicated, the returned weight matrix ({\tt e(omega)}) will store these weights with a final column providing the numerical ID of units, where this numerical ID is either taken from the unit variable (if this variable is a numerical format), or arranged in alphabetical order based on the unit variable, if this variable is in string format.

\subsubsection{Returned Objects}
\texttt{sdid} stores the following in \texttt{e()}:

Scalars: 

\begin{tabular}{ll}        
      \texttt{e(ATT)}       & Average Treatment Effect on the Treated  \\
      \texttt{e(se)}        & Standard error for the ATT  \\
      \texttt{e(reps)}      & Number of bootstrap/placebo replications \\
      \texttt{e(N\_clust)}  & Number of clusters \\
\end{tabular}

Macros: 

\begin{tabular}{ll}        
      \texttt{e(cmd)}       &     sdid  \\
      \texttt{e(cmdline)}   &     command as typed \\
      \texttt{e(depvar)}    &     name of dependent variable \\
      \texttt{e(vce)}       &     vcetype specified in \texttt{vce()} \\
      \texttt{e(clustvar)}  &     name of cluster variable\\
\end{tabular}

Matrices: 

\begin{tabular}{ll}  			
      \texttt{e(tau)}        &      tau estimator for each adoption time-period\\
      \texttt{e(lambda)}     &     lambda weights (time-specific weights)\\
      \texttt{e(omega)}      &     omega weights (unit-specific weights)\\
      \texttt{e(adoption)}   &     adoption times\\
      \texttt{e(beta)}     &      beta vector corresponding to covariates\\
      \texttt{e(series)}   &      control and treatment series of the graphs (only returned \\
    &                      when the graph option is indicated)\\
     \texttt{e(difference)}   & difference between treatment and control series (only returned \\
    &                      when the graph option is indicated)
\end{tabular} 

\vspace{3mm}
\noindent

\section{Examples based on an Empirical Application}
\label{scn:examples}
In the sections below we provide a number of illustrations of the usage of, and performance of, the {\tt sdid} command, which operationalizes the Synthetic Difference-in-Differences estimator in Stata.  We consider both a block treatment design (with a single adopting state), and a staggered adoption design, noting a number of points covering estimation, inference, and visualization.

\subsection{A Block Design}

\leftskip 0.1in
\parindent +0.1in

In the first case, we consider the well-known example, also presented in \citet{Arkhangelskyetal2021}, of California's ``Proposition 99'' tobacco control measure.  This example, based on the context described in \citet{Abadieetal2010} and data of \citet{OrzechowskiWalker2005}, is frequently used to illustrate synthetic control style methods.  Proposition 99, which was passed by California in 1989, increased the taxes paid on a packet of cigarettes by 25 cents.  The impact of this reform is sought to be estimated by comparing the evolution of sales of cigarettes in packs per capita in California (the treated state) with those in 38 untreated states, which did not significantly increase cigarette taxes during the study period.

The data used in analysis cover each of these 39 states over the period of 1970--2000, with a single observation for each state and year.  Adoption occurs in California in 1989, implying $T_{pre}=19$ pre-treatment periods and $T_{post}=12$ post-treatment periods.  There are $N_{co}=38$ control and a single treated state, hence $N_{tr}=1$.  Using the \texttt{sdid} command, we replicate the results from \citet{Arkhangelskyetal2021}.  In the below code example, we first download the data, and then conduct the Synthetic Difference-in-Differences implementation using a placebo inference procedure with (a default) 50 placebo iterations.

\begin{stlog}
    \input{examples/ex1_1.log}
\end{stlog}
The third line of this code excerpt quite simply implements the synthetic difference-in-differences estimator, returning identical point estimates to those documented in Table 1 of \citet{Arkhangelskyetal2021}.  Standard errors are slightly different, as these are based on pseudo-random placebo reshuffling, though can be replicated as presented here provided that the same seed is set in the {\tt seed} option.  Note that in this case, given that a small number (1) of treated units is present, placebo inference is the only appropriate procedure, as indicated in the {\tt vce()} option.

\begin{figure}[htpb!]
\caption{Proposition 99, example from \citet{Abadieetal2010,Arkhangelskyetal2021}}
\label{fig:sdidplot}
\subfloat[Unit-Specific Weights]{%
\includegraphics[width=0.5\textwidth]{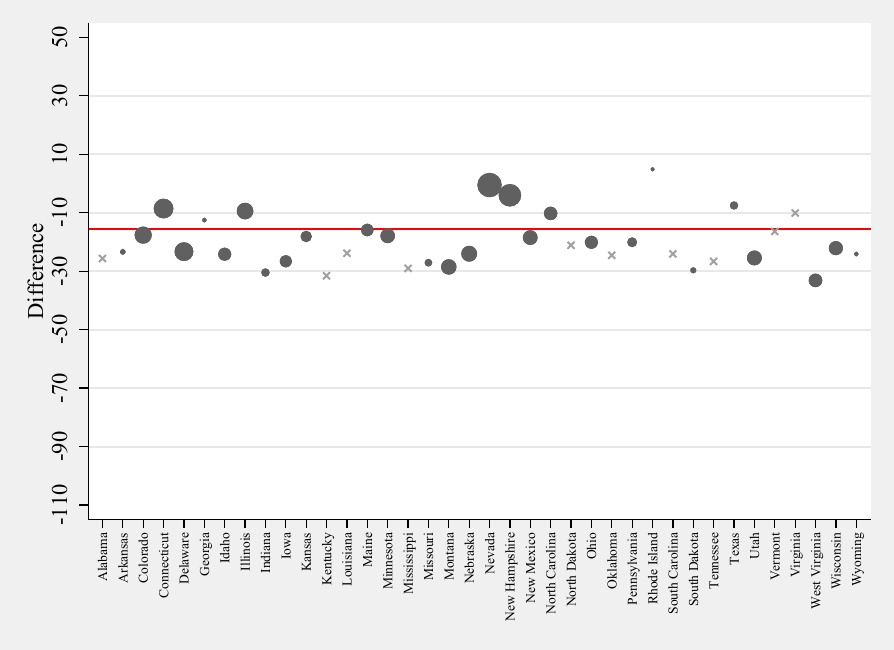}%
}
\subfloat[Outcome Trends and Time-Specific Weights]{%
 \includegraphics[width=0.5\textwidth]{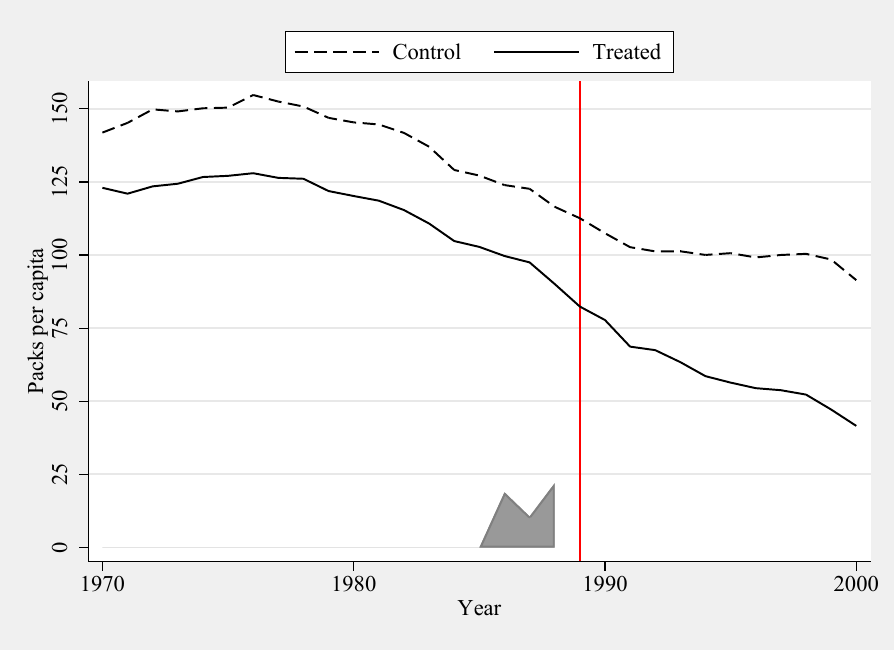}%
}
\end{figure}

Should we wish to generate the same graphs as in \citet{Arkhangelskyetal2021}, summarizing both (a) unit specific weights, and (b) treatment and synthetic control outcome trends along with time specific weights, this can be requested with the addition of the {\tt graph} option.  This is displayed below, where we additionally modify plot aesthetics via the {\tt g1\_opt()} and {\tt g2\_opt()} options for weight graphs (Figure \ref{fig:sdidplot}(a)), and trend graphs (Figure \ref{fig:sdidplot}(b)) respectively.  Finally, generated graphs can be saved to disk using the {\tt graph\_export()} option, with a graph type (\texttt{.eps} below), and optionally a pre-pended plot name.  Output corresponding to the below command is provided in Figure \ref{fig:sdidplot}.

\begin{stlog}
. sdid packspercapita state year treated, vce(placebo) seed(1213) graph g1on
>     g2_opt(ylabel(0(25)150) ytitle("Packs per capita") scheme(sj))
>     g1_opt(xtitle("") scheme(sj)) g1on graph_export(sdid_, .eps)
\end{stlog}


It is illustrative to compare the output of SDID estimation procedures with those of standard synthetic control methods of \citet{Abadieetal2010}, and unweighted difference-in-difference estimates.  By using the {\tt method()} option one can request a standard difference-in-differences output, requested with {\tt method(did)}, or synthetic control output, requested with {\tt method(sc)}. In the interests of completeness, {\tt method(sdid)} is also accepted, although this is the default behaviour when {\tt method} is not included in command syntax.  In each case, resulting graphs document matched treated and control/synthetic control trends, as well as weights received by each unit and time period.  These are displayed in Figure \ref{fig:californiaplot}, with plots corresponding to each of the three calls to {\tt sdid} displayed below. In the left-hand panel, identical SDID plots are provided as those noted above.  In the middle plot, corresponding to {\tt method(did)}, a difference-in-difference setting is displayed.  Here, in the top panel, outcomes for California are displayed as a solid line, while mean outcomes for all control states are documented as a dashed line, where a clear divergence is observed in the pre-treatment period.  The bottom panel shows that in this case, each control unit receives an identical weight, while time weights indicated at the base of the top plot note that each period is weighted identically.  Finally, in the case of synthetic control, output from the third call to {\tt sdid} is provided in the right-hand panel.  In this case, treated and synthetic control units are observed to overlap nearly exactly, with weights in figure (f) noted to be more sparse, and placing relatively more weight on fewer control states.  We note that in each case, the {\tt vce(noinference)} option is used, as here we are simply interested in observing exported graphs, not the entire command output displaying aggregate estimates, standard errors and confidence intervals. 

\begin{stlog}
    \input{examples/comparison.log}
\end{stlog}

\begin{figure}[htpb!]
\caption{Comparison of estimators}
\label{fig:californiaplot}
\subfloat[SDID: Outcome Trends]{%
 \includegraphics[width=0.33\textwidth]{./graphs/sdid_trends1989.pdf}%
}
\subfloat[DID: Outcome Trends]{%
 \includegraphics[width=0.33\textwidth]{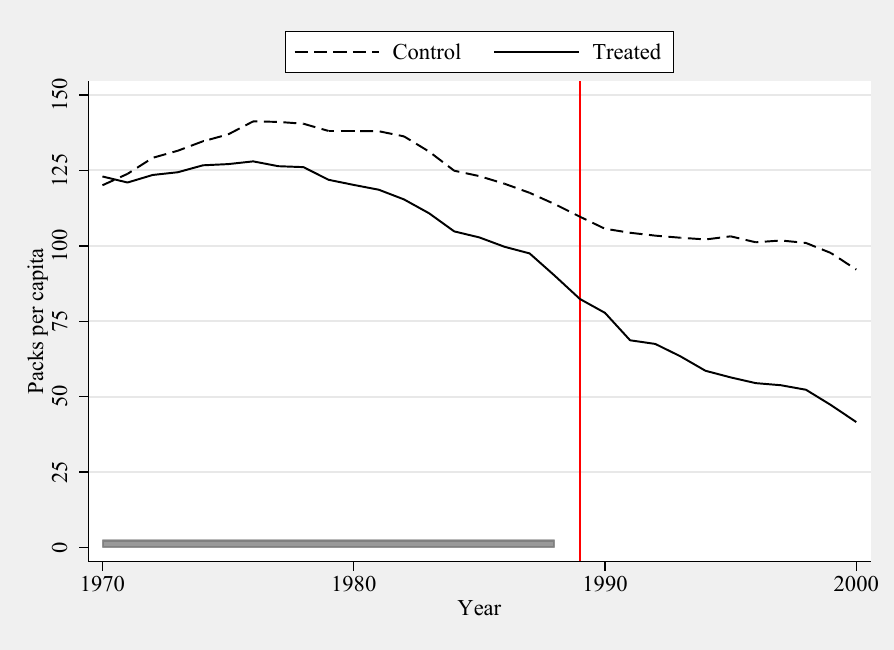}%
}
\subfloat[SC: Outcome Trends]{%
 \includegraphics[width=0.33\textwidth]{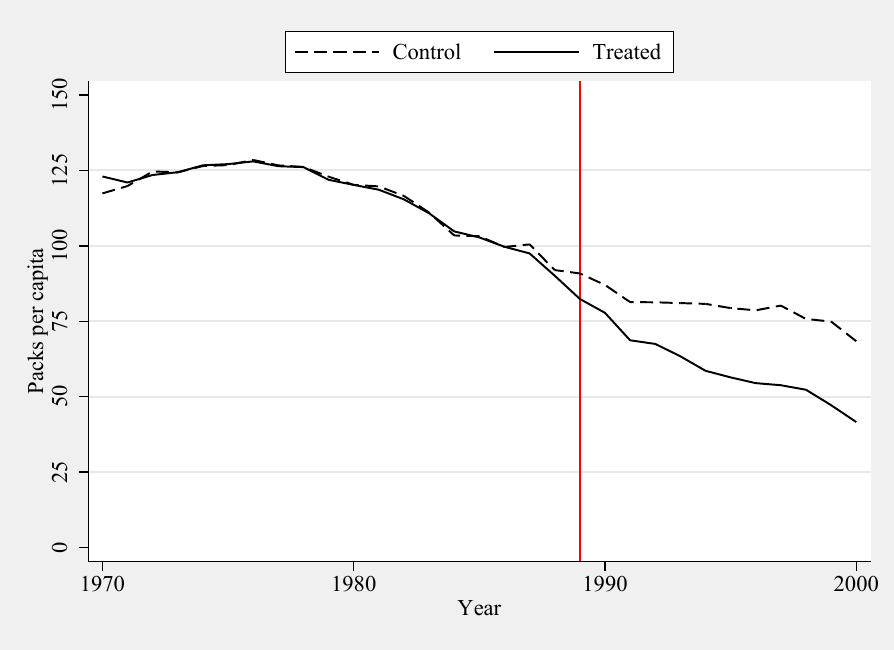}%
}\\
\subfloat[SDID: Unit-Specific Weights]{%
\includegraphics[width=0.33\textwidth]{./graphs/sdid_weights1989.pdf}%
}
\subfloat[DID: Unit-Specific Weights]{%
\includegraphics[width=0.33\textwidth]{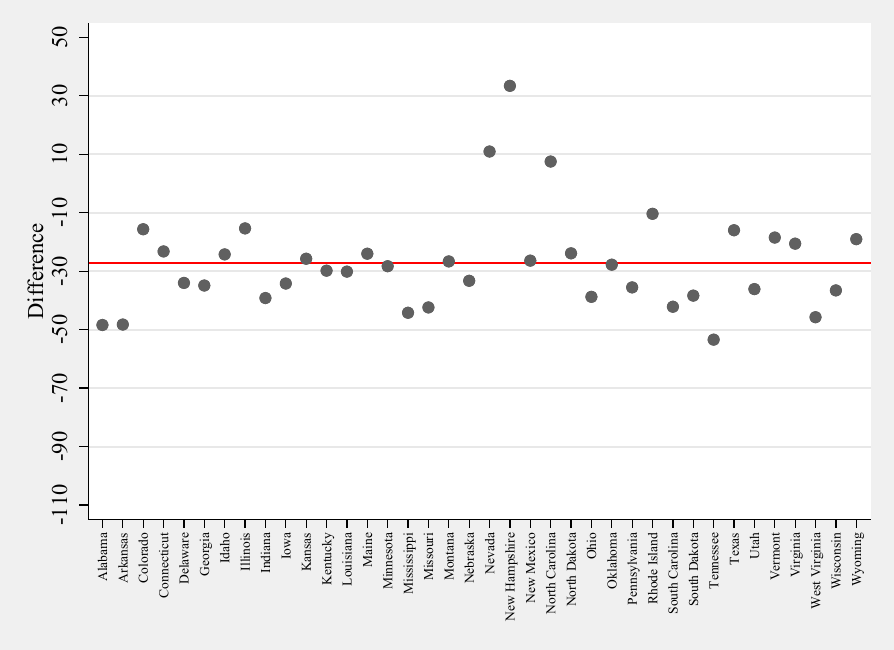}%
}
\subfloat[SC: Unit-Specific Weights]{%
\includegraphics[width=0.33\textwidth]{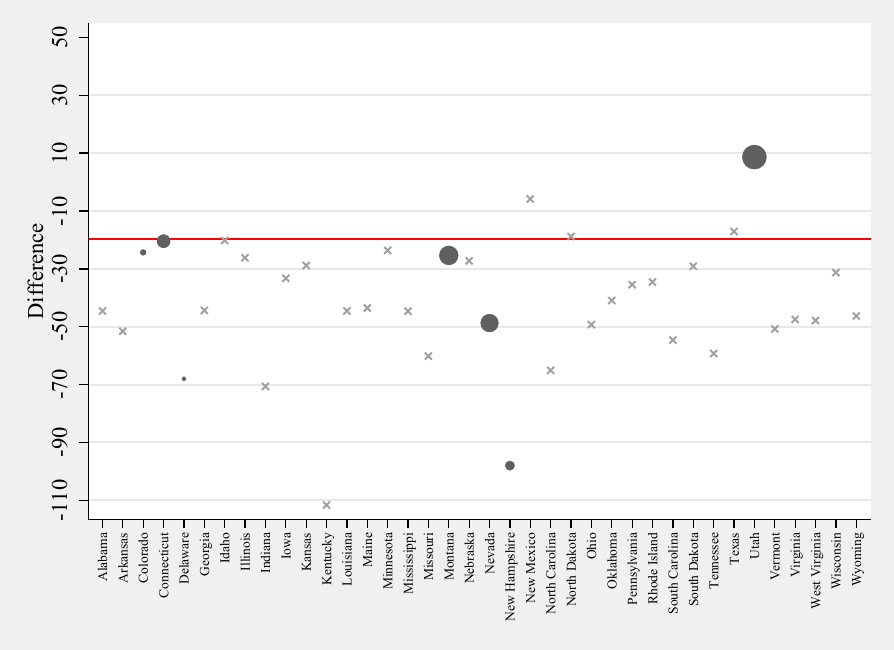}%
}
\end{figure}

\begin{figure}[htpb!]
\caption{Outcome Trends and Time-Specific Weights}
\label{fig:sdidplotmanual}
\includegraphics[width=0.9\textwidth]{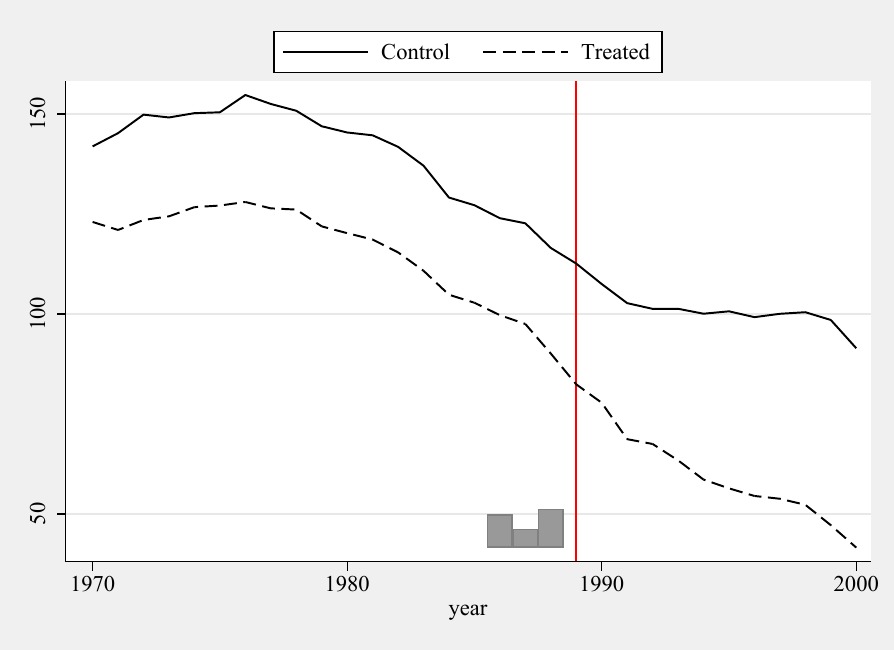}
\end{figure}

The {\tt sdid} command returns multiple matrices containing treatment and control outcome trends, weights, and other elements.  These elements can be accessed simply for use in post-estimation procedures or graphing.  As a simple example, the following code excerpt accesses treatment and synthetic control outcome trends (stored in {\tt e(series)}, and time weights (stored in {\tt e(lambda)}) and uses these elements to replicate the plot presented in Figure \ref{fig:sdidplot}b.  The resulting graphical output is presented  as Figure \ref{fig:sdidplotmanual}.  Thus, if one wishes to have further control over the precise nature of plotting, beyond that provided in the graphing options available in {\tt sdid}'s command syntax, one can simply work with elements returned in the ereturn list.  In Appendix \ref{app:additionalResults}, we show that with slightly more effort, returned elements can be used to construct the unit-specific weight plot from Figure \ref{fig:sdidplot}a.


\begin{stlog}
    \input{examples/ex2_1.log}
\end{stlog}

\subsection{A Staggered Adoption Design}
\label{sscn:SADexample}
We present an example of a staggered adoption design, based on data and the context studied in \citet{Bhalotraetal2022}.  In this case, the impact of parliamentary gender quotas which reserve seats for women in parliament are estimated, first on rates of women in parliament, and secondly on rates of maternal mortality.  This is conducted on a country by year panel, where for each of 1990-2015, 115 countries are observed, 9 of which implement a parliamentary gender quota.\footnote{This is a subset of the full sample studied in \citet{Bhalotraetal2022}.  Here we only work with countries for which observations of women in parliament and maternal mortality exist for the full time period, without missing observations.}  For each of these countries, data on the rates of women in parliament and the maternal mortality ratio are collected, as well as a number of covariates.  

This example presents a staggered adoption configuration, given that in the period under study, quota adoption occurred in seven different yearly periods between 2000 and 2013. {\tt sdid} handles a staggered adoption configuration seamlessly without any particular changes to the syntax.  In the code below, we implement the synthetic difference-in-differences estimator using the bootstrap procedure to calculate standard errors.  The output by default reports the weighted ATT which is defined in (\ref{eqn:ATT}) above.  However, as laid out in (\ref{eqn:ATT}), this is based on each adoption-period specific synthetic difference-in-differences estimate.  These adoption-period specific estimates are returned in the matrix {\tt e(tau)}, which is tabulated below the standard command output.

\begin{stlog}
    \input{examples/ex3_1.log}
\end{stlog}

All other elements are identical to those documented in the case of a single adoption period, however generalised to multiple adoptions.  For example, if requesting graphical output, a single treatment versus synthetic control trend graph and corresponding unit-level weight graph is provided for each adoption date.  Similarly, {\tt ereturn}ed matrices such as {\tt e(lambda)}, {\tt e(omega)} and {\tt e(series)} provide columns for each particular adoption period.

\paragraph{Adding Covariates}
As laid out in Section \ref{sscn:controls}, covariates can be handled in synthetic difference-in-differences in a number of ways.  Below we document the inclusion of a single covariate (the natural logarithm of GDP per capita).  As {\tt sdid} is based on a balanced panel of observations, we must first ensure that there are no missing observations for all covariates, in this case dropping a small number of (control) countries for which this measure is not available. We then include covariates via the {\tt covariates()} option.  In the first case, this is conducted exactly following the procedure discussed by \citet{Arkhangelskyetal2021}, in which parameters on covariates are estimated within the optimization routines in Mata.  This is analogous to indicating {\tt covariates(, optimized)}.  Estimates in this particular case suggest that the inclusion of this control does little to dampen effects.  After estimation, the coefficients on the covariates can be inspected as part of {\tt e(beta)}, where an adoption-specific value for each covariate is provided, given that the underlying SDID estimate is calculated for each adoption period.

\begin{stlog}
    \input{examples/ex4_1.log}
\end{stlog}

The inclusion of covariates in the previous implementation adds considerably to the computational time as it increases the complexity of the underlying optimization routine, and this is conducted in each adoption period and each bootstrap replicate.  An alternative manner to capture covariates described in section \ref{sscn:controls} above is that of \citet{Kranz2022}, where the impact of covariates are projected out using a baseline regression of the outcome on covariates and fixed effects only in units where the treatment status is equal to zero.  This is implemented as below, with {\tt covariates(, projected)}.

\begin{stlog}
    \input{examples/ex5_1.log}
\end{stlog}

Here, results are slightly different, though quantitatively comparable to those when using alternative procedures for conditioning out covariates.  In this case, if examining the {\tt e(beta)} matrix, only a single coefficient will be provided, as the regression used to estimate the coefficient vector is always based on the same sample.  This additionally offers a non-trivial speed up in the execution of the code.  For example, on a particular personal computer with Stata SE 15.1 and relatively standard specifications, using the {\tt optimized} method above requires 324 seconds of computational time while using {\tt projected} requires 61 seconds (compared with 58 seconds where covariates are not included in {\tt sdid}).

\paragraph{Post-Estimation Commands}
While {\tt sdid} provides a standard tabular and graphical output as displayed previously, the command can used to provide output in alternative formats.  For example, the {\tt sdid} command interacts seamlessly with routines such as {\tt estout} \citep{Jann2004} for the exportation of results tables.  To see this, the below block of code estimates three specific versions of the model discussed above, storing each model using an {\tt eststo:} prefix, before finally exporting estimated ATTs and standard errors to a LaTeX file, which can be included in tabular form as displayed in Table \ref{tab:regtab}.  Similar such procedures could be conducted with routines such as {\tt outreg} or {\tt outreg2}, and tabular output could be further enriched using additional options within {\tt esttab} if desired. 

%
\begin{stlog}
. webuse set www.damianclarke.net/stata/
. webuse quota_example.dta, clear
. lab var quota "Parliamentary Gender Quota"

. eststo sdid_1: sdid womparl country year quota, vce(bootstrap) seed(2022)

. drop if lngdp==.
. eststo sdid_2: sdid womparl country year quota, vce(bootstrap) seed(2022) 
>     covariates(lngdp, optimized)

. eststo sdid_3: sdid womparl country year quota, vce(bootstrap) seed(2022)
>     covariates(lngdp, projected)
\end{stlog}

\begin{stlog}
. esttab sdid_1 sdid_2 sdid_3 using "example1.tex", 
>     nonotes nomtitles stats(N, labels("Observations") fmt(\%9.0fc))
>     addnotes("* p$<$0.10, ** p$<$0.05, *** p$<$0.01")
>     starlevel ("*" 0.10 "**" 0.05 "***" 0.01) lab
>     b(\%-9.3f) se(\%-9.3f) style(tex) replace
\end{stlog}

\begin{table}[htpb!]
    \centering
    \caption{Tabular Output Following {\tt sdid}}
    \label{tab:regtab}
\input{examples/example1}
\end{table}

\subsection{Inference Options}
\label{sscn:inferenceEmp}
In this section we provide examples of the implementation of alternative inference options, as laid out in algorithms \ref{alg:BoostrapStaggered}-\ref{alg:placeboStaggered}.  For this illustration we will keep only treated units which adopt gender quotas in 2002 and 2003, as otherwise adoption periods will exist in which only a single unit is treated, and jackknife procedures will not be feasible.

\begin{stlog}
    \input{examples/ex6_1.log}
\end{stlog}

In the following three code blocks we document bootstrap, placebo and jackknife inference procedures.  The difference in implementation in each case is very minor, simply indicating either {\tt bootstrap}, {\tt placebo} or {\tt jaccknife} in the {\tt vce()} option.  For example, in the case of bootstrap inference, where block bootstraps over the variable {\tt country} are performed, the syntax is as follows:

\medskip
\begin{stlog}
    \input{examples/ex7_1.log}
\end{stlog}

By default, only 50 bootstrap replicates are performed, though in practice, a substantially higher number should be used, and this can be indicated in the {\tt reps(\#)} option.  In the case of placebo, the syntax and output are virtually identical.  The suitability of each method depends on the underlying structure of the panel, and in this particular case, given the relatively small number of treated units, it may be the case that placebo procedures are preferred. 

\begin{stlog}
    \input{examples/ex8_1.log}
\end{stlog}

Finally, in the interests of completeness, the jackknife procedure, which is by far the fastest of the three to execute\footnote{As an example, with 50 replicates for bootstrap and placebo, and on a standard personal computer running Stata SE, 15.1, the execution time for bootstrap was 18.2 seconds, for placebo permutations was 10.09 seconds, and for jackknife was 0.7 seconds.  This time scales approximately linearly with the number of replicates in the case of bootstrap and placebo.  With 500 replicates the time was 178.1 and 101.6 for bootstrap and placebo procedures respectively.}, is provided below.  Note that unlike the case with placebo or bootstrap inference, it is not necessary (or relevant) to set a seed, nor indicate the number of replications, as the jackknife procedure implies conducting a leave-one-out procedure over each unit.  In this particular case, jackknife inference appears to be more conservative than bootstrap procedures, in line with what may be expected based on \citet{Arkhangelskyetal2021}'s demonstration that jackknife inference is in general conservative. 

\begin{stlog}
    \input{examples/ex9_1.log}
\end{stlog}



\subsection{Event Study Style Output}
While {\tt sdid} offers a simple implementation to conduct standard synthetic difference-in-difference procedures and provide output, with some work results can also be visualized in alternative ways.  
For example, consider the standard `panel event-study' style setting (see \emph{e.g.} \citet{Freyaldenhovenetal2019,SchmidheinySiegloch2019,ClarkeTapia2021}), where one wishes to visualize how the dynamics of some treatment effect evolve over time, as well as how differences between treated and control units evolve prior to the adoption of treatment.  Such graphs are frequently used to efficiently provide information on both the credibility of parallel pre-trends in an observational setting, as well as the emergence of any impact owing to treatment once treatment is switched on.  

\begin{figure}[htpb!]
\caption{Outcome trends and event study style estimate of the impact of quotas on \% women in parliament}
\label{fig:eventstudymanual}
\subfloat[Outcome Trends]{%
 \includegraphics[width=0.5\textwidth]{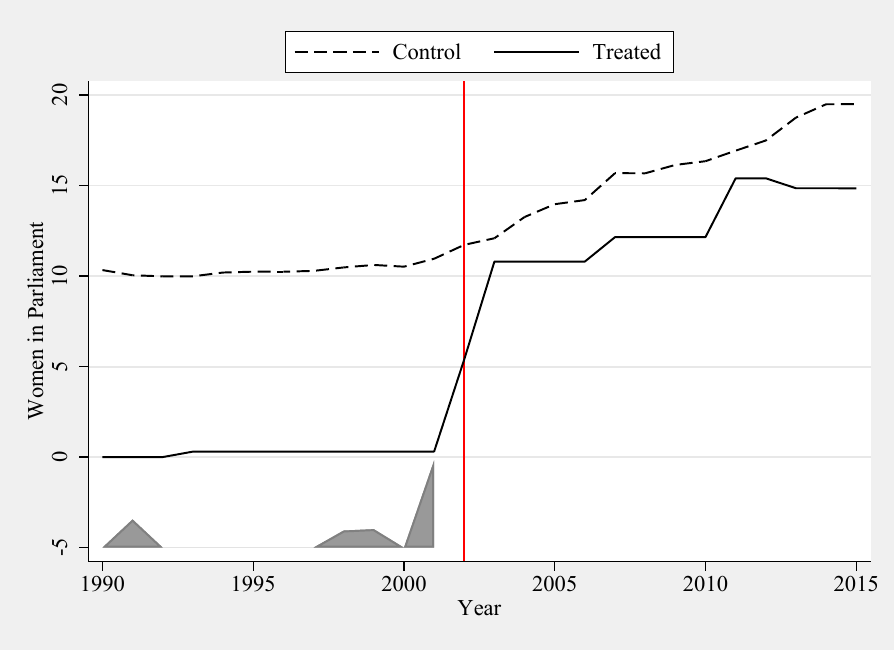}%
}
\subfloat[Event Study]{%
 \includegraphics[width=0.5\textwidth]{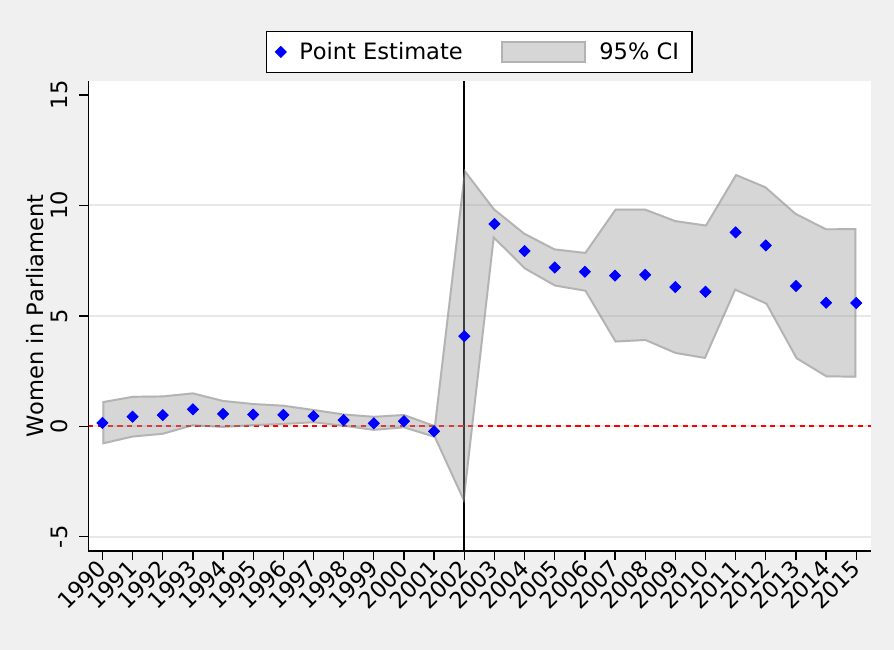}%
}
\end{figure}

What such an analysis seeks to document is the differential evolution of treated and (synthetic) control units, abstracting away from any baseline difference between the groups.  As an example, refer to Figure \ref{fig:eventstudymanual}(a), which is based on the adoption of gender quotas laid out in section \ref{sscn:SADexample}, and in particular quota adoption year 2002.  This is standard output from {\tt sdid}, presenting trends in rates of women in parliament in countries which adopted quotas in 2002 (solid line), and synthetic control countries which did not adopt quotas (dashed line).  We will refer to the values plotted in these trend lines as $\bar{Y}^{Tr}_t$ for treated units in year $t$, and $\bar{Y}^{Co}_t$ for synthetic control units in year $t$.   While this standard output allows us to visualize trends in the two groups in a simple way, it is not immediately clear how the \emph{differences} in these outcomes evolve over time compared to baseline differences, 
nor the confidence intervals on any such changes over time.  

For this to resemble the logic of an event study analysis, we wish to consider, for each period $t$, whether differences between treated units and synthetic controls have changed when compared to baseline differences.  Namely, for each period $t$, we wish to calculate:
\begin{equation}
\label{eqn:eventT}
 \left(\bar{Y}^{Tr}_t-\bar{Y}^{Co}_t\right)-\left(\bar{Y}^{Tr}_{baseline}-\bar{Y}^{Co}_{baseline}\right),
\end{equation}
along with the confidence interval for this quantity. Here $\bar{Y}^{Tr}_{baseline}$ and $\bar{Y}^{Co}_{baseline}$ refer to baseline (pre-treatment) means for treated and synthetic control units respectively.  In standard panel event studies, some arbitrary baseline period is chosen off of which to estimate pre-treatment differences.  This is often one year prior to treatment.  In the case of SDID where pre-treatment weights are optimally chosen as $\widehat\lambda^{sdid}_t$ (refer to section \ref{scn:methods}), this suggests an alternative quantity for $\bar{Y}^{Tr}_{baseline}$ and $\bar{Y}^{Co}_{baseline}$, namely: 
\begin{equation}
\label{eqn:baseline}
\bar{Y}^{Tr}_{baseline} = \sum_{t=1}^{T_{pre}}\widehat\lambda^{sdid}_t\bar{Y}^{Tr}_t \qquad \bar{Y}^{Co}_{baseline} = \sum_{t=1}^{T_{pre}}\widehat\lambda^{sdid}_t\bar{Y}^{Co}_t.
\end{equation}
In words these baseline outcomes are simply pre-treatment aggregates, where weights are determined by optimal pre-treatment weights (indicated by the shaded gray area in Figure \ref{fig:eventstudymanual}(a)).  The event study then plots the quantities defined in (\ref{eqn:eventT}), for each time $t$.


An example of such an event study style plot is presented in Figure \ref{fig:eventstudymanual}(b).  Here, blue points present the quantity indicated in (\ref{eqn:eventT}) for each year.  In this case, $t$ ranges from 1990 to 2015.  
While all these points are based off a simple implementation of {\tt sdid} comparing outcomes between treated and control units following (\ref{eqn:eventT}), confidence intervals documented in gray shaded areas of Figure \ref{fig:eventstudymanual}(b) can be generated following the resampling or permutation procedures discussed earlier in this paper.   Specifically, in the case of re-sampling, a block bootstrap can be conducted, and in each iteration the quantity in (\ref{eqn:eventT}) can be re-calculated for each $t$.  The confidence interval associated with each of these quantities can then be calculated based on its variance across many (block)-bootstrap resamples.



Figure \ref{fig:eventstudymanual}(b), and graphs following this principle more generally, can be generated following the use of {\tt sdid}.  However, by default {\tt sdid} simply provides output on trends among the treated and synthetic control units (as displayed in Figure  \ref{fig:eventstudymanual}(a)). In the code below, we lay out how one can move from these trends to the event study in panel (b).  As this procedure requires conducting the inference portion of the plot manually (unlike most other procedures involving {\tt sdid} where inference is conducted automatically as part of the command) the code is somewhat more involved.  For this reason, we discuss the code below in a number of blocks, terminating with the generation of the plot displayed in Figure \ref{fig:eventstudymanual}(b).

In a first code block, we will open the parliamentary gender quota data which we used in section \ref{sscn:SADexample}, and keep the particular adoption period considered here (countries which adopt quotas in 2002), as well as un-treated units:

\begin{stlog}
. webuse set www.damianclarke.net/stata/
. webuse quota_example.dta, clear

. egen m=min(year) if quota==1, by(country)  //indicator for the year of adoption
. egen mm=mean(m), by(country)
. keep if mm==2002 | mm==.                   //keep only one time of adoption
. drop if lngdp==.                           //keep if covariates observed
\end{stlog}

We can then implement the standard SDID procedure, additionally exporting the trend graphs which is displayed in Figure \ref{fig:eventstudymanual}(a).  This is done in the first line below, after which a number of vectors are stored.  These vectors allow us to calculate the quantity $(\bar{Y}^{Tr}_{baseline}-\bar{Y}^{Co}_{baseline})$ indicated in (\ref{eqn:eventT}), which is generated from $\widehat\lambda^{sdid}$, from the returned matrix {\tt e(lambda)}, and pre-treatment values for $\bar{Y}^{Tr}_t$ and $\bar{Y}^{Co}_t$, from the returned matrix {\tt e(series)}.  This baseline quantity is referred to as {\tt meanpre\_o} below.  Finally, the quantity of interest in (\ref{eqn:eventT}) for each time period $t$ is generated as the variable {\tt d}, which is plotted below as the blue points on the event study in Figure \ref{fig:eventstudymanual}(b).


\begin{stlog}
. qui sdid womparl country year quota, vce(noinference) graph g2_opt(ylab(-5(5)20) 
>     ytitle("Women in Parliament") scheme(sj)) graph_export(groups, .pdf) 
>     covariates(lngdp, projected)

. matrix lambda = e(lambda)[1..12,1]    //save lambda weight
. matrix yco = e(series)[1..12,2]       //control baseline
. matrix ytr = e(series)[1..12,3]       //treated baseline
. matrix aux = lambda'*(ytr - yco)      //calculate the pre-treatment mean
. scalar meanpre_o = aux[1,1]    

. matrix difference = e(difference)[1..26,1..2]  // Store Ytr-Yco
. svmat difference
. ren (difference1 difference2) (time d)
. replace d = d - meanpre_o                      // Calculate vector in (8)
\end{stlog}

Perhaps the most complicated portion of code is that which implements the bootstrap procedure.  This is provided below, where for ease of replication we consider a relatively small number of bootstrap resamples, which are set as the local {\tt B = 100}.  In each bootstrap resample, we first ensure that both treatment and control units are present (using the locals {\tt r1} and {\tt r2}), and then re-estimate the {\tt sdid} procedure with the new bootstrap sample generated using Stata's {\tt bsample} command.  This is precisely the same block bootstrap procedure laid out by \citet{Arkhangelskyetal2021}, and which {\tt sdid} conducts internally, however here we are interested in collecting, for each bootstrap resample, the same quantity estimated above with the main sample as {\tt d}, which captures the estimate defined in (\ref{eqn:eventT}) for each $t$.  To do so, we simply follow an identical procedure as that conducted above, however now save the resulting resampled values of the quantities from (\ref{eqn:eventT}) as a series of matrices \verb|d`b'| for later processing to generate confidence intervals in the event study.




\begin{stlog}
. local b = 1
. local B = 100   
. while `b'<=`B' \{
.     preserve
.     bsample, cluster(country) idcluster(c2)
.     qui count if quota == 0
.     local r1 = r(N)
.     qui count if quota != 0
.     local r2 = r(N)    
.     if (`r1'!=0 & `r2'!=0) \{
.         qui sdid womparl c2 year quota, vce(noinference) graph covariates(lngdp, projected)
		
.         matrix lambda_b = e(lambda)[1..12,1]       //save lambda weight
.         matrix yco_b = e(series)[1..12,2]          //control baseline
.         matrix ytr_b = e(series)[1..12,3]          //treated baseline
.         matrix aux_b = lambda_b'*(ytr_b - yco_b)  //calculate the pre-treatment mean
.         matrix meanpre_b = J(26,1,aux_b[1,1])
		
.         matrix d`b' = e(difference)[1..26,2] - meanpre_b
		
.         local ++b
.     \}
.     restore
. \}
\end{stlog}

%

The final step is to calculate the standard deviation of each estimate from (\ref{eqn:eventT}) based on the bootstrap resamples, and then to generate confidence intervals for each parameter based on the estimates generated above ({\tt d}), as well as their standard errors.  This is conducted in the first lines of the code below.  For each of the $B=100$ resamples conducted above, we import the vector of resampled estimates from (\ref{eqn:eventT}), and then using {\tt rowsd()} calculate the standard deviation of the estimates across each time period $t$.  This is the bootstrap standard error, which is used below to calculate the upper and lower bounds of 95\% confidence intervals as [{\tt LCI};{\tt UCI}].  Finally, based on these generated elements ({\tt d}, as blue points on the event study, and {\tt LCI, UCI} as the end points of confidence intervals) we generate the output for Figure \ref{fig:eventstudymanual}(b) in the final lines of code.

\begin{stlog}
. preserve
. keep time d
. keep if time!=.

. forval b=1/`B' \{
.     svmat d`b'   // import each bootstrap replicate of difference between trends
. \}

. egen rsd = rowsd(d11 - d`B'1)      //calculate standard deviation of this difference
. gen LCI = d + invnormal(0.025)*rsd //lower bounds on bootstrap CIs
. gen UCI = d + invnormal(0.975)*rsd //upper bounds on bootstrap CIs

. *generate plot
. tw rarea UCI LCI time, color(gray
>     xtitle("") ytitle("Women in Parliament") xlab(1990(1)2015, angle(45))
>     legend(order(2 "Point Estimate" 1 "95
>     xline(2002, lc(black) lp(solid)) yline(0, lc(red) lp(shortdash)) 
>     scheme(sj)
. graph export "event_sdid.pdf", replace
. restore
\end{stlog}
As noted above, the outcome of this graph is provided in Figure \ref{fig:eventstudymanual}(b), where we observe that, as expected, the synthetic difference-in-difference algorithm has resulted in quite closely matched trends between the synthetic control and treatment group in the pre-treatment period, given that all pre-treatment estimates lie close to zero.  The observed impact of quotas on women in parliament occurs from the treatment year onward, where these differences are observed to be large and statistically significant.

This process of estimating an event study style plot is conducted here for a specific adoption year.  In the case of a block adoption design where there is only one adoption period, this will be the only resulting event study to consider.  However in a staggered adoption design, a single event study could be generated for each adoption period.  Potentially, such event studies could be combined, but some way would be required to deal with unbalanced lags and leads, and additionally some weighting function would be required to group treatment lags and leads where multiple such lags and leads are available.  One such procedure has been proposed in \citet{AbrahamSun2018}, and could be a way forward here.

\section{Conclusions}
In this paper we have laid out the details behind \citet{Arkhangelskyetal2021}'s SDID method, and discussed its implementation in Stata using the {\tt sdid} command.  We have briefly discussed the methods behind this command, as well as laid out extensions into a staggered adoption setting.  We provide two empirical examples to demonstrate the usage of the command.

It is important to note that given the nature of the algorithm, a number of requirements must be met for this to be applied to data. We lay these out below, as key considerations for empirical researchers wishing to conduct estimation and inference using the SDID estimator.
\begin{itemize}
    \item Firstly, and most importantly, a balanced panel of data is required that provides outcomes and treatment measures for each unit in all periods under study.  Should missing values in such outcomes be present in the panel, these either must be eliminated from the estimation sample, or data should be sought to fill in gaps.  
    \item Secondly, no units can be considered if they were exposed to treatment from the first period in which data is observed.  If this occurs, there is no pre-treatment period on which to generate synthetic control cohorts.  If always treated units are present in the data, these either need to be eliminated, or earlier data sought.
    \item Third, pure control units are required.  At least some units must never be treated in order to act as donor units.  If all units are treated at some point in the panel, no donor units exist, and synthetic controls cannot be generated.
    \item Fourth, in cases where covariates are included, these covariates must be present in all observations.  If missing observations are present in covariates, this will generate similar problems as when outcomes or treatment measures are missing.  If missing observations are present, these treated units shold be removed from the estimation sample, or data should be sought to complete the covariate coverage.
    \item Finally, in the case of inference, a number of additional requirements must be met.  In the case of bootstrap or jackknife procedures, the number of treated units should be larger than 1 (and ideally considerably larger than this).  Should only 1 treated unit be present, placebo inference should be conducted.  Additionally, in the case of placebo inference, this can only be conducted if the number of control units exceeds the number of treated units.
\end{itemize}

Should a balanced panel of data be available, the SDID method, and the {\tt sdid} command described here, offers a flexible, easy to implement and robust option for the analysis of impacts of policies or treatments in certain groups at certain times.  These methods provide clear graphical results to describe outcomes, and an explicit description of how counterfactual outcomes are inferred.  These methods are likely well suited to a large body of empirical work in social sciences, where treatment assignment is not random, and offer benefits over both difference-in-differences and synthetic control methods.

\newpage
\bibliography{sj}

\begin{aboutauthors}
Susan Athey is the Economics of Technology Professor at Stanford Graduate School of Business.

Damian Clarke is an Associate Professor at The Department of Economics of The Universidad de Chile, a Research Fellow at IZA and an Associate at the Millennium Institute for Market Imperfections and Public Policy and CAGE, Warwick.  

Guido Imbens is the Applied Econometrics Professor and Professor of Economics at Stanford Graduate School of Business.

Daniel Pailañir is an MA student at The Department of Economics of The Universidad de Chile, and a young researcher associated with the Millennium Nucleus for the Study of the Life Course and Vulnerability.
\end{aboutauthors}

\begin{Athanks}
We are grateful to Asjad Naqvi for comments relating to this code, and many users of the {\tt sdid} ado for sending feedback and suggestions related to certain features implemented here.
\end{Athanks}

\newpage
\appendix
\section*{Appendices}
\section{Estimation Algorithms for the Block Design}
\label{app:algorithms}
\setcounter{algocf}{0}
\renewcommand{\thealgocf}{A\arabic{algocf}}
In this appendix, we replicate the estimation algorithm and inference algorithms defined in \citet{Arkhangelskyetal2021}.  These are referred to in the text, and follow the same notation as in section \ref{scn:methods} here.

\begin{algorithm}
\caption{Algorithm 1 from \citet{Arkhangelskyetal2021}}
\label{alg:alg1}
Data: \textbf{Y}, \textbf{W}. \\
Result: Point estimate $\widehat{\tau}^{sdid}$.  \\ \ \\

1. Compute regularization parameter $\zeta$\;
2. Compute unit weights $\hat\omega^{sdid}$\;
3. Compute time weights $\hat\lambda^{sdid}$\;
4. Compute the SDID estimator via the weighted DID regression\;
\[\left(\hat\tau^{sdid},\hat\mu,\hat\alpha,\hat\beta\right)=\argmin_{\tau,\mu,\alpha,\beta}\left\{\sum_{i=1}^N\sum_{t=1}^T(Y_{it}-\mu-\alpha_i-\beta_t-W_{it}\tau)^2\widehat\omega^{sdid}_i\hat\lambda^{sdid}_t \right\}\]
\end{algorithm}

\begin{algorithm}
\caption{Algorithm 2 from \citet{Arkhangelskyetal2021}}
Data: \textbf{Y}, \textbf{W}, $B$. \\
Result: Variance estimator $\widehat{V}_{\tau}^{cb}$ \\ \ \\
\For{b $\leftarrow$ 1 to B}
{
1. Construct a bootstrap dataset $(\mathbf{Y}^{(b)},\mathbf{W}^{(b)})$ by sampling $N$ rows of $(\mathbf{Y},\mathbf{W})$ with replacement\;
2. \If{the bootstrap sample has no treated units or no control units}
{
\hspace{0.2cm}Discard resample and go to 1\;
}\hspace{0.4cm}\textbf{end}\\ 
3. Compute SDID estimate $\tau^{(b)}$ following algorithm \ref{alg:alg1} based on $(\mathbf{Y}^{(b)},\mathbf{W}^{(b)})$\;
}\textbf{end}\\ \ \\
4. Define $\widehat{V}_{\tau}^{cb}=\frac{1}{B}\sum_{b=1}^B\left(\widehat{\tau^{(b)}}-\frac{1}{B}\sum_{b=1}^B\widehat{\tau^{(b)}}\right)^2$;
\end{algorithm}

\begin{algorithm}
\caption{Algorithm 3 from \citet{Arkhangelskyetal2021}}
Data: $\widehat\omega$, $\widehat\lambda$ \textbf{Y}, \textbf{W}, $\widehat{\tau}$. \\
Result: Variance estimator $\widehat{V}_{\tau}$ \\ \ \\
\For{i $\leftarrow$ 1 to N}
{
1. Compute $\widehat{\tau}^{(-i)}: \argmin_{\tau,\{\alpha_j,\beta_t\}_{j\neq i,t}}\sum_{j\neq i,t}(Y_{it}-\alpha_j-\beta_t-W_{it}\tau)^2\widehat\omega_j\hat\lambda_t$\;
}\textbf{end}\\ \ \\
2. Compute $\widehat{V}_\tau^{jack}=(N-1)N^{-1}\sum_{i=1}^N\left(\widehat{\tau}^{(-i)}-\widehat{\tau}\right)^2$;
\end{algorithm}

\begin{algorithm}
\caption{Algorithm 4 from \citet{Arkhangelskyetal2021}}
Data: $\mathbf{Y}_{co}$, $N_{tr}$, $B$. \\
Result: Variance estimator $\widehat{V}_{\tau}^{placebo}$ \\ \ \\
\For{b $\leftarrow$ 1 to B}
{
1. Sample $N_{tr}$ out of the $N_{co}$ control units without replacment to `receive the placebo'\;
2. Construct a placebo treatment matrix $\mathbf{W}^{(b)}_{co}$, for the controls\; 
3. Compute SDID estimate $\tau^{(b)}$ based on $(\mathbf{Y}_{co},\mathbf{W}^{(b)}_{co})$\;
}\textbf{end}\\ \ \\
4. Define $\widehat{V}_{\tau}^{placebo}=\frac{1}{B}\sum_{b=1}^B\left(\widehat{\tau^{(b)}}-\frac{1}{B}\sum_{b=1}^B\widehat{\tau^{(b)}}\right)^2$;
\end{algorithm}
\clearpage

\section{Replicating Weight Graphs}
\label{app:additionalResults}
After implementing the {\tt sdid} estimator, the unit specific weights can be used to re-create the weight graph provided as output automatically with the {\tt graph} option.  While this is somewhat involved, and likely would not be conducted by hand, it may be illustrative to see how this is generated, combining both unit-specific weights, and unit-specific difference-in-difference estimates. This code is displayed below, first saving time weights which are used to calculate DID estimates, secondly saving unit weights for graphing, thirdly combining all elements and calculating the DID estimates, and  finally, generating the graph, which is displayed after this code excerpt.

\begin{stlog}
\input{examples/ex2_2_1.log}
\end{stlog}

\setcounter{figure}{0}
\renewcommand{\thefigure}{A\arabic{figure}}
\begin{figure}[htpb!]
\includegraphics[width=0.99\textwidth]{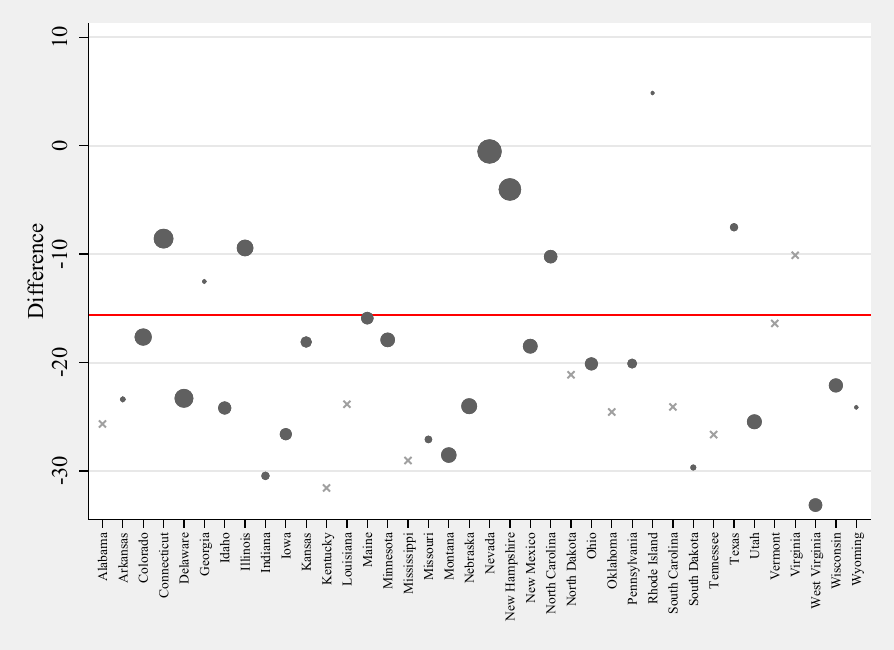}
\caption{Unit-Specific Weight Graph}
\label{fig:sdidplotmanual2}
\end{figure}

\end{document}

%% file: examples/ex1_1.log.tex
. webuse set www.damianclarke.net/stata/
{\smallskip}
. webuse prop99_example.dta, clear
{\smallskip}
. sdid packspercapita state year treated, vce(placebo) seed(1213)
Placebo replications (50). This may take some time.
----+--- 1 ---+--- 2 ---+--- 3 ---+--- 4 ---+--- 5
..................................................     50
{\smallskip}
{\smallskip}
Synthetic Difference-in-Differences Estimator
{\smallskip}
\HLI{13}{\TOPT}\HLI{63}
packsperca{\tytilde}a {\VBAR}     ATT     Std. Err.     t      P>|t|    [95\% Conf. Interval]
\HLI{13}{\PLUS}\HLI{63}
     treated {\VBAR} -15.60383    9.53183    -1.64    0.102   -34.28588     3.07822
\HLI{13}{\BOTT}\HLI{63}
95\% CIs and p-values are based on Large-Sample approximations.
Refer to Arkhangelsky et al., (2020) for theoretical derivations.
{\smallskip}

%% file: examples/comparison.log.tex
sdid packspercapita state year treated, method(sdid) vce(noinference) graph
>     g1_opt(ylabel(-110(20)50) xtitle("") scheme(sj)) g1on
>     g2_opt(ylabel(0(25)150) ytitle("Packs per capita") scheme(sj))
>     graph_export(sdid_, .eps)

sdid packspercapita state year treated, method(did) vce(noinference) graph msize(small)
>     g1_opt(ylabel(-110(20)50) xtitle("") scheme(sj)) g1on
>     g2_opt(ylabel(0(25)150) ytitle("Packs per capita") scheme(sj))
>     graph_export(did_, .eps)

sdid packspercapita state year treated, method(sc) vce(noinference) graph 
>     g1_opt(ylabel(-110(20)50) xtitle("") scheme(sj)) g1on
>     g2_opt(ylabel(0(25)150) ytitle("Packs per capita") scheme(sj))
>     graph_export(sc_, .eps)

%% file: examples/ex2_1.log.tex
. preserve
. clear
. matrix series=e(series)
. matrix lambda=e(lambda)
. qui svmat series, names(col)
. qui svmat lambda

. tw line Yco1989 year, yaxis(1) || 
>    line Ytr1989 year, yaxis(1) || 
>    bar lambda1 year if year<=1988, yaxis(2) ylabel(0(1)5, axis(2)) yscale(off axis(2))
>    xline(1989, lc(red)) legend(order(1 "Control" 2 "Treated") pos(12) col(2)) scheme(sj)
{\smallskip}
. graph export sdid\_replicate.eps, replace
{\smallskip}
. restore
{\smallskip}

%% file: examples/ex3_1.log.tex
. webuse quota_example.dta, clear
{\smallskip}
. sdid womparl country year quota, vce(bootstrap) seed(1213)
Bootstrap replications (50). This may take some time.
----+--- 1 ---+--- 2 ---+--- 3 ---+--- 4 ---+--- 5
..................................................     50
{\smallskip}
{\smallskip}
Synthetic Difference-in-Differences Estimator
{\smallskip}
\HLI{13}{\TOPT}\HLI{63}
     womparl {\VBAR}     ATT     Std. Err.     t      P>|t|    [95\% Conf. Interval]
\HLI{13}{\PLUS}\HLI{63}
       quota {\VBAR}   8.03410    3.74040     2.15    0.032     0.70305    15.36516
\HLI{13}{\BOTT}\HLI{63}
95\% CIs and p-values are based on Large-Sample approximations.
Refer to Arkhangelsky et al., (2020) for theoretical derivations.
{\smallskip}
. matlist e(tau)
{\smallskip}
             {\VBAR}       Tau       Time 
\HLI{13}{\PLUS}\HLI{22}
          r1 {\VBAR}  8.388868       2000 
          r2 {\VBAR}  6.967746       2002 
          r3 {\VBAR}  13.95226       2003 
          r4 {\VBAR} -3.450543       2005 
          r5 {\VBAR}  2.749036       2010 
          r6 {\VBAR}  21.76272       2012 
          r7 {\VBAR} -.8203235       2013 
{\smallskip}

%% file: examples/ex4_1.log.tex
. drop if lngdp==.
{\smallskip}
. sdid womparl country year quota, vce(bootstrap) seed(1213) covariates(lngdp)
Bootstrap replications (50). This may take some time.
----+--- 1 ---+--- 2 ---+--- 3 ---+--- 4 ---+--- 5
..................................................     50
{\smallskip}
{\smallskip}
Synthetic Difference-in-Differences Estimator
{\smallskip}
\HLI{13}{\TOPT}\HLI{63}
     womparl {\VBAR}     ATT     Std. Err.     t      P>|t|    [95\% Conf. Interval]
\HLI{13}{\PLUS}\HLI{63}
       quota {\VBAR}   8.05150    3.09252     2.60    0.009     1.99027    14.11272
\HLI{13}{\BOTT}\HLI{63}
95\% CIs and p-values are based on Large-Sample approximations.
Refer to Arkhangelsky et al., (2020) for theoretical derivations.
{\smallskip}

%% file: examples/ex5_1.log.tex
. sdid womparl country year quota, vce(bootstrap) seed(1213) covariates(lngdp, projected)
Bootstrap replications (50). This may take some time.
----+--- 1 ---+--- 2 ---+--- 3 ---+--- 4 ---+--- 5
..................................................     50
{\smallskip}
{\smallskip}
Synthetic Difference-in-Differences Estimator
{\smallskip}
\HLI{13}{\TOPT}\HLI{63}
     womparl {\VBAR}     ATT     Std. Err.     t      P>|t|    [95\% Conf. Interval]
\HLI{13}{\PLUS}\HLI{63}
       quota {\VBAR}   8.05927    3.11913     2.58    0.010     1.94589    14.17264
\HLI{13}{\BOTT}\HLI{63}
95\% CIs and p-values are based on Large-Sample approximations.
Refer to Arkhangelsky et al., (2020) for theoretical derivations.
{\smallskip}

%% file: examples/example1.tex
{
\def\sym#1{\ifmmode^{#1}\else\(^{#1}\)\fi}
\begin{tabular}{l*{3}{c}}
\hline\hline
            &\multicolumn{1}{c}{(1)}   &\multicolumn{1}{c}{(2)}   &\multicolumn{1}{c}{(3)}   \\
\hline
Parliamentary Gender Quota&       8.034** &       8.051***&       8.059***\\
            &     (3.940)   &     (3.047)   &     (3.099)   \\
\hline
Observations&       3,094   &       2,990   &       2,990   \\
\hline\hline
\multicolumn{4}{l}{\footnotesize * p$<$0.10, ** p$<$0.05, *** p$<$0.01}\\
\end{tabular}
}

%% file: examples/ex6_1.log.tex
. webuse quota\_example.dta, clear
. drop if country=="Algeria"   | country=="Kenya" | country=="Samoa" | 
>         country=="Swaziland" | country=="Tanzania"

%% file: examples/ex7_1.log.tex
. sdid womparl country year quota, vce(bootstrap) seed(1213)
Bootstrap replications (50). This may take some time.
----+--- 1 ---+--- 2 ---+--- 3 ---+--- 4 ---+--- 5
..................................................     50
{\smallskip}
{\smallskip}
Synthetic Difference-in-Differences Estimator
{\smallskip}
\HLI{13}{\TOPT}\HLI{63}
     womparl {\VBAR}     ATT     Std. Err.     t      P>|t|    [95\% Conf. Interval]
\HLI{13}{\PLUS}\HLI{63}
       quota {\VBAR}  10.33066    4.72911     2.18    0.029     1.06178    19.59954
\HLI{13}{\BOTT}\HLI{63}
95\% CIs and p-values are based on Large-Sample approximations.
Refer to Arkhangelsky et al., (2020) for theoretical derivations.
{\smallskip}

%% file: examples/ex8_1.log.tex
. sdid womparl country year quota, vce(placebo) seed(1213)
Placebo replications (50). This may take some time.
----+--- 1 ---+--- 2 ---+--- 3 ---+--- 4 ---+--- 5
..................................................     50
{\smallskip}
{\smallskip}
Synthetic Difference-in-Differences Estimator
{\smallskip}
\HLI{13}{\TOPT}\HLI{63}
     womparl {\VBAR}     ATT     Std. Err.     t      P>|t|    [95\% Conf. Interval]
\HLI{13}{\PLUS}\HLI{63}
       quota {\VBAR}  10.33066    5.14741     2.01    0.045     0.24191    20.41941
\HLI{13}{\BOTT}\HLI{63}
95\% CIs and p-values are based on Large-Sample approximations.
Refer to Arkhangelsky et al., (2020) for theoretical derivations.
{\smallskip}

%% file: examples/ex9_1.log.tex
. sdid womparl country year quota, vce(jackknife)                   
{\smallskip}
{\smallskip}
Synthetic Difference-in-Differences Estimator
{\smallskip}
\HLI{13}{\TOPT}\HLI{63}
     womparl {\VBAR}     ATT     Std. Err.     t      P>|t|    [95\% Conf. Interval]
\HLI{13}{\PLUS}\HLI{63}
       quota {\VBAR}  10.33066    6.00560     1.72    0.085    -1.44009    22.10141
\HLI{13}{\BOTT}\HLI{63}
95\% CIs and p-values are based on Large-Sample approximations.
Refer to Arkhangelsky et al., (2020) for theoretical derivations.
{\smallskip}

%% file: examples/ex2_2_1.log.tex
. preserve
. clear
. matrix lambda=e(lambda)
. svmat lambda
. ren (lambda1 lambda2) (lambda year)
. keep if year<=1988
. tempfile dlambda
. save `dlambda'
. restore

. preserve
. clear
. matrix omega=e(omega)
. svmat omega
. ren (omega1 omega2) (omega id)
. keep if id<=39
. tempfile domega
. save `domega'
. restore

. merge m:1 year using `dlambda', nogen
. bys state: egen y1=mean(packspercapita) if year>=1989
. bys state: egen y2=mean(packspercapita*lambda) if year<=1988
. replace y2=y2*19
. egen ypost=mean(y1), by(state)
. egen ypre=mean(y2), by(state)
. keep state ypre ypost
. duplicates drop
. gen delta=ypost-ypre
. qui sum delta if state=="California"
. gen sdelta=`r(mean)'
. gen difference=sdelta-delta
. egen id=group(state)
. merge 1:1 id using `domega', nogen
. drop if omega==.
. encode state, gen(state2) l(id)

. tw scatter difference state2 if omega!=0 [aw=omega], msize(tiny) || 
>     scatter difference state2 if omega==0, m(X) 
>     xlabel(1(1)38, angle(vertical) labs(vsmall) valuelabel)
>     yline(-15.60383, lc(red)) legend(off) xtitle("")
>    ytitle("Difference") scheme(sj)
. graph export sdid\_panela.eps, replace

%% file: sdidStataJournal.bbl
\ifnum 31=1 \def\bibname{Reference}
\else \def\bibname{References} \fi
\begin{thebibliography}{31}
\expandafter\ifx\csname natexlab\endcsname\relax\def\natexlab#1{#1}\fi
\expandafter\ifx\csname url\endcsname\relax
  \def\url#1{\texttt{#1}}\fi
\expandafter\ifx\csname urlprefix\endcsname\relax\def\urlprefix{URL }\fi

\bibitem[{Abadie et~al.(2010)Abadie, Diamond, and Hainmueller}]{Abadieetal2010}
Abadie, A., A.~Diamond, and J.~Hainmueller. 2010.
\newblock {Synthetic Control Methods for Comparative Case Studies: Estimating
  the Effect of California’s Tobacco Control Program}.
\newblock \emph{Journal of the American Statistical Association} 105(490):
  493--505.

\bibitem[{Abadie et~al.(2015)Abadie, Diamond, and Hainmueller}]{Abadieetal2015}
---------. 2015.
\newblock Comparative Politics and the Synthetic Control Method.
\newblock \emph{American Journal of Political Science} 59(2): 495--510.
\urlprefix\url{https://onlinelibrary.wiley.com/doi/abs/10.1111/ajps.12116.}
\bibitem[{Abadie and Gardeazabal(2003)}]{AbadieGardeazabal2003}
Abadie, A., and J.~Gardeazabal. 2003.
\newblock {The Economic Costs of Conflict: A Case Study of the Basque Country}.
\newblock \emph{American Economic Review} 93(1): 113--132.

\bibitem[{Abadie and L'Hour(2021)}]{AbadieLHour2021}
Abadie, A., and J.~L'Hour. 2021.
\newblock A Penalized Synthetic Control Estimator for Disaggregated Data.
\newblock \emph{Journal of the American Statistical Association} 116(536):
  1817--1834.

\bibitem[{Arkhangelsky et~al.(2021)Arkhangelsky, Athey, Hirshberg, Imbens, and
  Wager}]{Arkhangelskyetal2021}
Arkhangelsky, D., S.~Athey, D.~A. Hirshberg, G.~W. Imbens, and S.~Wager. 2021.
\newblock Synthetic Difference-in-Differences.
\newblock \emph{American Economic Review} 111(12): 4088--4118.
\urlprefix\url{https://www.aeaweb.org/articles?id=10.1257/aer.20190159.}
\bibitem[{Athey and Imbens(2022)}]{AtheyImbens2022}
Athey, S., and G.~W. Imbens. 2022.
\newblock {Design-based analysis in Difference-In-Differences settings with
  staggered adoption}.
\newblock \emph{Journal of Econometrics} 226(1): 62--79.
\newblock Annals Issue in Honor of Gary Chamberlain.
\urlprefix\url{https://www.sciencedirect.com/science/article/pii/S0304407621000488.}
\bibitem[{Ben-Michael et~al.(2021)Ben-Michael, Feller, and
  Rothstein}]{BenMichealetal2021}
Ben-Michael, E., A.~Feller, and J.~Rothstein. 2021.
\newblock The Augmented Synthetic Control Method.
\newblock \emph{Journal of the American Statistical Association} 116(536):
  1789--1803.

\bibitem[{Bhalotra et~al.(2022)Bhalotra, Clarke, Gomes, and
  Venkataramani}]{Bhalotraetal2022}
Bhalotra, S.~R., D.~Clarke, J.~F. Gomes, and A.~Venkataramani. 2022.
\newblock {Maternal Mortality and Women’s Political Power}.
\newblock Working Paper 30103, National Bureau of Economic Research.
\urlprefix\url{http://www.nber.org/papers/w30103.}
\bibitem[{Bhuller et~al.(2013)Bhuller, Havnes, Leuven, and
  Mogstad}]{Bhulleretal2013}
Bhuller, M., T.~Havnes, E.~Leuven, and M.~Mogstad. 2013.
\newblock Broadband Internet: An Information Superhighway to Sex Crime?
\newblock \emph{Review of Economic Studies} 80(4): 1237--1266.
\urlprefix\url{https://EconPapers.repec.org/RePEc:oup:restud:v:80:y:2013:i:4:p:1237-1266.}
\bibitem[{{Bilinski} and {Hatfield}(2018)}]{BilinskiHatfield2018}
{Bilinski}, A., and L.~A. {Hatfield}. 2018.
\newblock {Nothing to see here? Non-inferiority approaches to parallel trends
  and other model assumptions}.
\newblock \emph{arXiv e-prints}  arXiv:1805.03273.

\bibitem[{de~Chaisemartin and D’Haultfœuille(2022)}]{dCDH2022}
de~Chaisemartin, C., and X.~D’Haultfœuille. 2022.
\newblock {Two-way fixed effects and differences-in-differences with
  heterogeneous treatment effects: a survey}.
\newblock \emph{The Econometrics Journal} Utac017.
\urlprefix\url{https://doi.org/10.1093/ectj/utac017.}
\bibitem[{Clarke and Tapia-Schythe(2021)}]{ClarkeTapia2021}
Clarke, D., and K.~Tapia-Schythe. 2021.
\newblock Implementing the panel event study.
\newblock \emph{The Stata Journal} 21(4): 853--884.
\urlprefix\url{https://doi.org/10.1177/1536867X211063144.}
\bibitem[{Conley and Taber(2011)}]{ConleyTaber2011}
Conley, T.~G., and C.~R. Taber. 2011.
\newblock Inference with “Difference in Differences” with a Small Number of
  Policy Changes.
\newblock \emph{The Review of Economics and Statistics} 93(1): 113--125.

\bibitem[{Doudchenko and Imbens(2016)}]{DoudchenkoImbens2016}
Doudchenko, N., and G.~W. Imbens. 2016.
\newblock Balancing, Regression, Difference-In-Differences and Synthetic
  Control Methods: A Synthesis.
\urlprefix\url{https://arxiv.org/abs/1610.07748.}
\bibitem[{Ferman and Pinto(2021)}]{FermanPinto2021}
Ferman, B., and C.~Pinto. 2021.
\newblock Synthetic controls with imperfect pretreatment fit.
\newblock \emph{Quantitative Economics} 12(4): 1197--1221.
\urlprefix\url{https://onlinelibrary.wiley.com/doi/abs/10.3982/QE1596.}
\bibitem[{Frank and Wolfe(1956)}]{FrankWolfe1956}
Frank, M., and P.~Wolfe. 1956.
\newblock An algorithm for quadratic programming.
\newblock \emph{Naval Research Logistics Quarterly} 3(1-2): 95--110.
\urlprefix\url{https://onlinelibrary.wiley.com/doi/abs/10.1002/nav.3800030109.}
\bibitem[{Freyaldenhoven et~al.(2019)Freyaldenhoven, Hansen, and
  Shapiro}]{Freyaldenhovenetal2019}
Freyaldenhoven, S., C.~Hansen, and J.~M. Shapiro. 2019.
\newblock Pre-event Trends in the Panel Event-Study Design.
\newblock \emph{American Economic Review} 109(9): 3307--38.
\urlprefix\url{http://www.aeaweb.org/articles?id=10.1257/aer.20180609.}
\bibitem[{Goodman-Bacon(2021)}]{GoodmanBacon2021}
Goodman-Bacon, A. 2021.
\newblock The Long-Run Effects of Childhood Insurance Coverage: Medicaid
  Implementation, Adult Health, and Labor Market Outcomes.
\newblock \emph{American Economic Review} 111(8): 2550--93.
\urlprefix\url{https://www.aeaweb.org/articles?id=10.1257/aer.20171671.}
\bibitem[{Hirshberg(undated)}]{HirshbergND}
Hirshberg, D.~A. Undated.
\newblock synthdid: Synthetic Difference in Differences Estimation.
\urlprefix\url{https://synth-inference.github.io/synthdid/.}
\bibitem[{Holland(1986)}]{Holand1986}
Holland, P.~W. 1986.
\newblock Statistics and Causal Inference.
\newblock \emph{Journal of the American Statistical Association} 81(396):
  945--960.

\bibitem[{Jann(2004)}]{Jann2004}
Jann, B. 2004.
\newblock {ESTOUT: Stata module to make regression tables}.
\newblock Statistical Software Components, Boston College Department of
  Economics.
\urlprefix\url{https://ideas.repec.org/c/boc/bocode/s439301.html.}
\bibitem[{Kranz(2022)}]{Kranz2022}
Kranz, S. 2022.
\newblock {Synthetic Difference-in-Differences with Time-Varying Covariates}.
\newblock Technical report.
\urlprefix\url{Available online at:
  \href{https://github.com/skranz/xsynthdid/blob/main/paper/synthdid_with_covariates.pdf}{https://github.com/skranz/xsynthdid/blob/main/paper/synthdid\_with\_covariates.pdf}.}
\bibitem[{Lawphongpanich(2009)}]{Lawphongpanich2009}
Lawphongpanich, S. 2009.
\newblock \emph{Encyclopedia of Optimization}, chap. Frank--Wolfe Algorithm,
  1094--1097.
\newblock Boston, MA: Springer US.

\bibitem[{Manski and Pepper(2018)}]{ManskiPepper2018}
Manski, C.~F., and J.~V. Pepper. 2018.
\newblock {How Do Right-to-Carry Laws Affect Crime Rates? Coping with Ambiguity
  Using Bounded-Variation Assumptions}.
\newblock \emph{The Review of Economics and Statistics} 100(2): 232--244.
\urlprefix\url{https://doi.org/10.1162/REST\_a\_00689.}
\bibitem[{Orzechowski and Walker(2005)}]{OrzechowskiWalker2005}
Orzechowski, W., and R.~C. Walker. 2005.
\newblock {The Tax Burden on Tobacco}.
\newblock Historical Compilation Volume 40, Arlington, VA.

\bibitem[{Paila{\~n}ir and Clarke(2022)}]{PailanirClarke2022}
Paila{\~n}ir, D., and D.~Clarke. 2022.
\newblock {SDID: Stata module to perform synthetic difference-in-differences
  estimation, inference, and visualization}.
\newblock Statistical Software Components, Boston College Department of
  Economics.
\urlprefix\url{https://ideas.repec.org/c/boc/bocode/s459058.html.}
\bibitem[{Rambachan and Roth(2019)}]{RambachanRoth2019}
Rambachan, A., and J.~Roth. 2019.
\newblock An Honest Approach to Parallel Trends.

\bibitem[{Roth et~al.(2022)Roth, Sant'Anna, Bilinski, and Poe}]{Rothetal2022}
Roth, J., P.~H.~C. Sant'Anna, A.~Bilinski, and J.~Poe. 2022.
\newblock What's Trending in Difference-in-Differences? A Synthesis of the
  Recent Econometrics Literature.
\urlprefix\url{https://arxiv.org/abs/2201.01194.}
\bibitem[{Rubin(2005)}]{Rubin2005}
Rubin, D.~B. 2005.
\newblock Causal Inference Using Potential Outcomes.
\newblock \emph{Journal of the American Statistical Association} 100(469):
  322--331.

\bibitem[{Schmidheiny and Siegloch(2019)}]{SchmidheinySiegloch2019}
Schmidheiny, K., and S.~Siegloch. 2019.
\newblock On Event Study Designs and Distributed-Lag Models: Equivalence,
  Generalization and Practical Implications.
\newblock IZA Discussion Papers 12079, Institute of Labor Economics (IZA).

\bibitem[{Sun and Abraham(2021)}]{AbrahamSun2018}
Sun, L., and S.~Abraham. 2021.
\newblock Estimating dynamic treatment effects in event studies with
  heterogeneous treatment effects.
\newblock \emph{Journal of Econometrics} 225(2): 175--199.
\urlprefix\url{https://EconPapers.repec.org/RePEc:eee:econom:v:225:y:2021:i:2:p:175-199.}
\end{thebibliography}
